\documentclass{aa}  
\usepackage{natbib}
\usepackage{graphicx}
\usepackage{txfonts}
\usepackage{float}
\usepackage{dblfloatfix}

\usepackage{multirow}
\usepackage{url}
 
\usepackage[colorlinks=true, linkcolor=blue, citecolor=blue, urlcolor=blue]{hyperref}

\usepackage{amsmath}

\usepackage{etoolbox}

\begin{document}

\title{Local constraints on alcohol--thioalcohol analogs in G35.2N\thanks{Corresponding authors: \\ Xuefang Xu, \email{xuefang\_xu@cqu.edu.cn} \\
Shanghuo Li, \email{shli@nju.edu.cn}\\
Qian Gou, \email{qian.gou@cqu.edu.cn}}}

\author{Xuefang Xu\inst{1,2}, Shanghuo Li\inst{3,4}, Qian Gou\inst{1,2}, Chunguo Duan\inst{1,2}, Laurent Pagani\inst{5}, Di Li\inst{6,7}, Jun Kang\inst{1}, Jiaxin Du\inst{1}, \and Jiaxiang Jiao\inst{1}}  
 
\institute{School of Chemistry and Chemical Engineering, Chongqing University, Chongqing 401331, China  
\and Chongqing Key Laboratory of Chemical Theory and Mechanism, Chongqing University, Chongqing 401331, China   
\and School of Astronomy and Space Science, Nanjing University, Nanjing, China
\and Key Laboratory of Modern Astronomy and Astrophysics, Nanjing University, Ministry of Education, Nanjing, China  
\and LUX, Observatoire de Paris, PSL Research University, CNRS, Sorbonne Universités, 75014 Paris, France
\and New Cornerstone Science Laboratory, Department of Astronomy, Tsinghua University, Beijing 100084, China 
\and National Astronomical Observatories, Chinese Academy of Sciences, Beijing 100012, China  
}

\date{}
 
\abstract
{The chemistry of sulfur-bearing complex organic molecules in dense
star-forming environments remains uncertain, partly because the dominant
sulfur reservoirs in dense gas and ices are poorly identified.
Alcohol--thioalcohol pairs provide a direct basis for comparing
the abundance behavior of structurally related O- and S-bearing
molecules. We present ALMA Band~6 observations of G35.2N and
investigate the alcohol--thioalcohol pairs CH$_3$OH/CH$_3$SH and
C$_2$H$_5$OH/C$_2$H$_5$SH toward three selected
spectral-extraction positions, MM3-pos, MM4-pos, and MM5-pos,
in the MM3--MM5 region. Local thermodynamic equilibrium spectral
modeling was used to derive molecular column densities and abundance
ratios. The CH$_3$OH column density was derived from
$^{13}$CH$_3$OH by adopting $^{12}$C/$^{13}$C = 50, because the main
isotopologue is affected by optical-depth effects.
Robust column-density constraints were obtained for CH$_3$OH,
CH$_3$SH, and C$_2$H$_5$OH at all three positions.
C$_2$H$_5$SH was robustly constrained toward MM3-pos and
MM5-pos but remains tentative toward MM4-pos. 
The adopted column densities vary moderately across the three
positions, with ranges of
$(1.8$--$2.7) \times 10^{18}$~cm$^{-2}$ for CH$_3$OH,
$(1.7$--$2.6) \times 10^{16}$~cm$^{-2}$ for CH$_3$SH,
$(5.7$--$8.5) \times 10^{16}$~cm$^{-2}$ for C$_2$H$_5$OH, and
$(2.4$--$4.6) \times 10^{15}$~cm$^{-2}$ for C$_2$H$_5$SH. 
Within the adopted uncertainties, the
CH$_3$OH/C$_2$H$_5$OH and CH$_3$OH/CH$_3$SH ratios are
consistent across MM3-pos, MM4-pos, and MM5-pos, with nominal
values of 31--32 and 100--110, respectively. 
Comparison with other chemically rich sources and warm-up
chemical models shows that the CH$_3$OH/C$_2$H$_5$OH ratio in
G35.2N lies within the range measured in other sources, whereas
CH$_3$OH/CH$_3$SH shows a larger source-to-source variation. The
ethyl-level O/S comparison remains less certain because many
literature C$_2$H$_5$SH measurements provide only lower limits. 
CH$_3$OH/CH$_3$SH is therefore the best-constrained O/S
alcohol--thioalcohol analog ratio in the present data and provides a
useful empirical probe of source-dependent sulfur-bearing organic
chemistry, while higher-sensitivity C$_2$H$_5$SH observations are
needed to test the C$_2$H$_5$OH/C$_2$H$_5$SH ratio.
}

\keywords{Astrochemistry --- Line: identification --- ISM: clouds --- ISM: molecules --- ISM: abundances}
\titlerunning{Local constraints on alcohol--thioalcohol analogs in G35.2N} \authorrunning{Xu et al.}\maketitle

\section{Introduction} 
\label{sec:intro}

Sulfur chemistry in star-forming environments remains poorly constrained. Observed gas-phase sulfur-bearing species account for only a small fraction of the expected elemental sulfur budget, implying that a substantial sulfur reservoir is hidden in grains, ices, or refractory material~\citep{Laas2019}. This sulfur-depletion problem complicates the interpretation of sulfur-bearing complex organic molecules, whose abundances may depend on both local physical conditions and the availability of reactive sulfur. Comparisons between S-bearing molecules and their O-bearing analogs provide a way to examine abundance differences within chemically related molecular families~\citep{Ceccarelli2017}.

The alcohol/thioalcohol pairs methanol/methanethiol and ethanol/ethanethiol are particularly suitable for this purpose. The two pairs, CH$_3$OH/CH$_3$SH and C$_2$H$_5$OH/C$_2$H$_5$SH, allow oxygen-to-sulfur substitution to be examined in closely related alkyl backbones. These pairs therefore provide direct O-to-S analog ratios within the alcohol/thioalcohol family, while constraining possible sulfur-specific effects. CH$_3$SH has been detected in both low- and high-mass star-forming environments, and alkanols and alkanethiols have been compared in chemically rich sources \citep{Majumdar2016,Muller2016,Gorai2021}. Laboratory and theoretical studies also indicate that sulfur-bearing organic molecules can form through grain-surface and ice-mediated pathways that need not mirror the corresponding O-bearing networks one-to-one~\citep{Lamberts2018,Santos2024}.

Previous observations have demonstrated the diagnostic value of
abundance ratios within the alcohol/thioalcohol family. In particular,
\citet{Rodriguez2021} compared SH-bearing molecules with their OH
analogs toward the Galactic-center cloud G+0.693 and placed these
ratios in the context of Sgr~B2(N2) and Orion~KL. 
Such studies have shown that ratios involving CH$_3$OH,
CH$_3$SH, C$_2$H$_5$OH, and C$_2$H$_5$SH can be used to
quantify abundance differences between thioalcohols and their
alcohol analogs. However, these comparisons
have largely relied on source-averaged or single-position
measurements. It therefore remains unclear whether the abundance
relationships among alcohol--thioalcohol analogs are
locally stable within a single high-mass star-forming source. 
A source-internal comparison separates abundance variations driven by 
source-to-source differences from those measured within 
a common physical and observational environment.
High-angular-resolution Atacama Large Millimeter/submillimeter
Array (ALMA) observations allow source-averaged abundance ratios to 
be complemented by measurements at selected chemically rich positions 
within a single source.

G35.2N, also known as G35.20$-$0.74N, is a high-mass star-forming region at a distance of 2.19~kpc and with a luminosity of $3\times10^4$~$L_\odot$ \citep{Dent1989,Sanchez2013}. Interferometric studies have resolved multiple compact continuum components and dense structures in this source \citep{Sanchez2014,Zhang2022}, together with disk-like kinematics and jet/outflow activity \citep{Sanchez2013,Beltran2016,Zhang2022}. Ordered magnetic-field morphologies have also been reported on comparable scales \citep{Qiu2013,Zhang2025,Hwang2026}. The chemical richness and compact continuum substructure of G35.2N provide the observational basis for a source-internal test of O/S alcohol--thioalcohol analog abundance ratios.

In this paper, we use ALMA Band~6 observations of G35.2N to compare
the alcohol/thioalcohol analogs CH$_3$OH, CH$_3$SH, C$_2$H$_5$OH,
and C$_2$H$_5$SH toward three selected spectral-extraction positions,
MM3-pos, MM4-pos, and MM5-pos, in
the MM3--MM5 region. The comparison is designed around two related
questions. First, does the purely O-bearing alcohol ratio
CH$_3$OH/C$_2$H$_5$OH remain stable among these positions? Second,
does the methyl-level O/S alcohol--thioalcohol analog ratio CH$_3$OH/CH$_3$SH show
similar source-internal stability, and how does it compare with ratios
measured in other chemically rich environments? The primary observational test is therefore
based on CH$_3$OH, CH$_3$SH, and C$_2$H$_5$OH, while C$_2$H$_5$SH
extends the comparison to the ethyl-level thioalcohol analog. 
The observations and data reduction are described in Sect.~\ref{sec:obs}.
The results are presented in Sect.~\ref{sec:res}, followed by the
discussion in Sect.~\ref{sec:dis} and the summary in Sect.~\ref{sec:sum}.

\section{Observations and data reduction}
\label{sec:obs}

\subsection{Observations}

The observations of G35.2N were obtained as part of the
Revealing the Origin of Multiplicity in high-mass protostellar
clusters with ALMA (ROMA) survey (Project ID:
2021.1.00713.S; PI: S. Li;
\citealt{Li2025}). The source was observed with
ALMA in Band~6, at a wavelength of 1.3 mm, using the 12 m
array in the C-5 and C-8 configurations. In this work, the 1.3 mm continuum image from the ROMA survey is used only as a spatial reference for identifying compact
structures, whereas the quantitative molecular-line analysis is
performed with the C-5 spectral-line cubes. We therefore
describe below the C-5 observational setup relevant to the
spectral analysis; details of the full ROMA observations are
given by~\citet{Li2025}.

The C-5 observations were carried out on 2022 July 5 with
41 antennas and baseline lengths of 15--1301 m. The correlator
setup included four spectral windows centered at 217.806,
220.577, 232.194, and 234.314 GHz. Each spectral window has a
bandwidth of 1.875 GHz and a channel spacing of 976.6 kHz,
corresponding to a velocity resolution of approximately
1.3 km s$^{-1}$. The total on-source integration time for
G35.2N in the C-5 observations was 1.5 min. The reference
phase center was
$(\alpha_{\rm ICRS}, \delta_{\rm ICRS}) =
(18^{\rm h}58^{\rm m}12\fs955,
+01\degr40\arcmin37\farcs367)$.

\begin{figure}[!htp]
  \centering
  \includegraphics[width=0.40\textwidth]{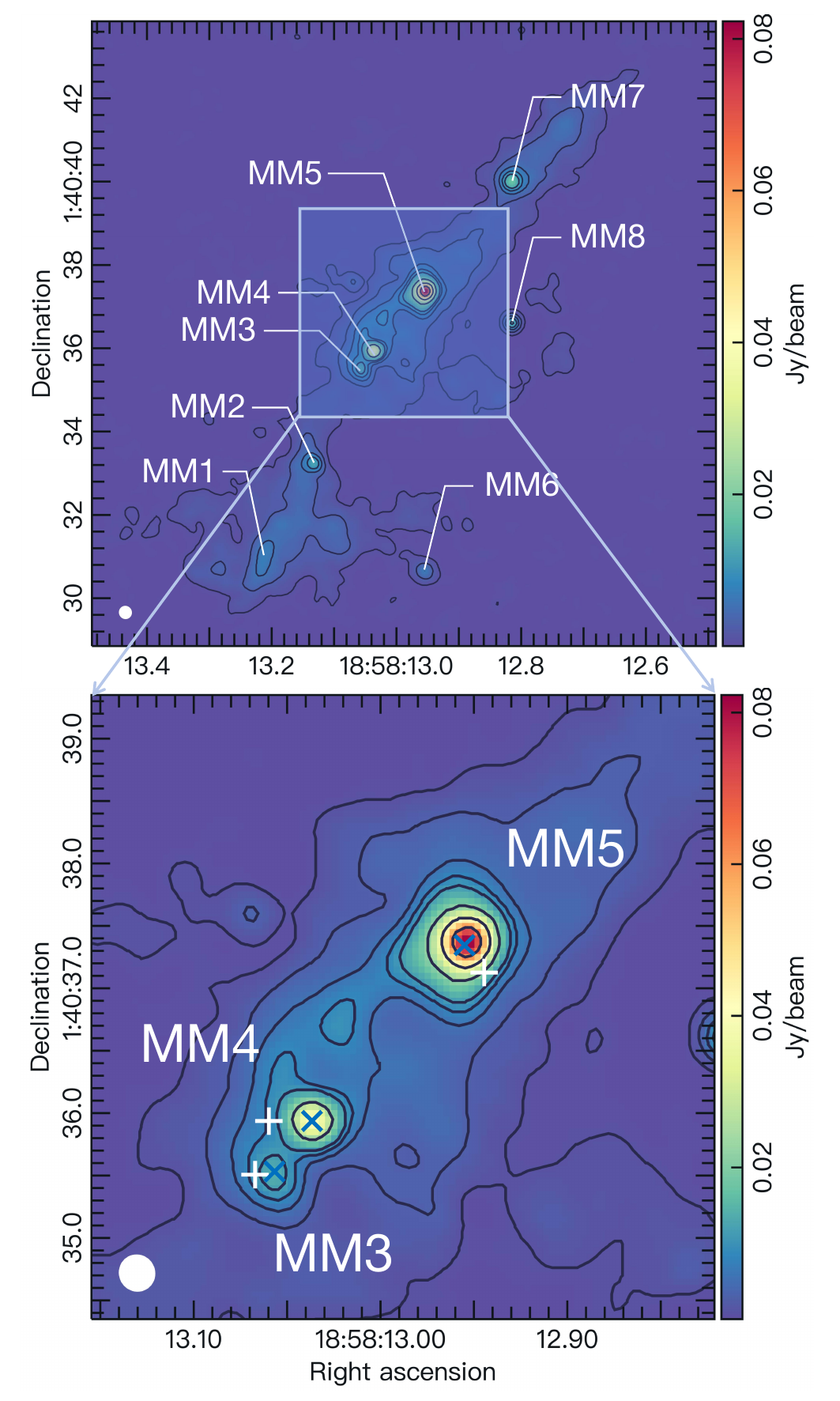}
  \caption{
(a) 1.3 mm continuum image of G35.2N obtained as part of the
ROMA survey. This continuum image is used only as a spatial
reference for the compact structures in the molecular-line
analysis. The contour levels are
$(3, 10, 20, 30, 40, 80, 140, 210) \times \sigma$, where
$\sigma = 0.27$\,mJy\,beam$^{-1}$ is the rms noise level of the
continuum image. The white ellipse in the lower-left corner
indicates the synthesized beam of the continuum image,
$0.29^{\prime\prime} \times 0.28^{\prime\prime}$, with a position
angle of $29.0\degr$.
(b) Zoomed-in view of the MM3, MM4, and MM5 region. 
The source labels MM3, MM4, and MM5 identify the corresponding
1.3 mm continuum structures. The ``$+$'' symbols mark the chemically
rich spectral-extraction positions adopted for the LTE analysis,
hereafter MM3-pos, MM4-pos, and MM5-pos, whereas the
``$\times$'' symbols mark the corresponding continuum peaks,
hereafter MM3-peak, MM4-peak, and MM5-peak.
}
\label{fig:cont}
\end{figure}

\begin{figure}[!htp]
  \centering
   \includegraphics[width=0.50\textwidth]{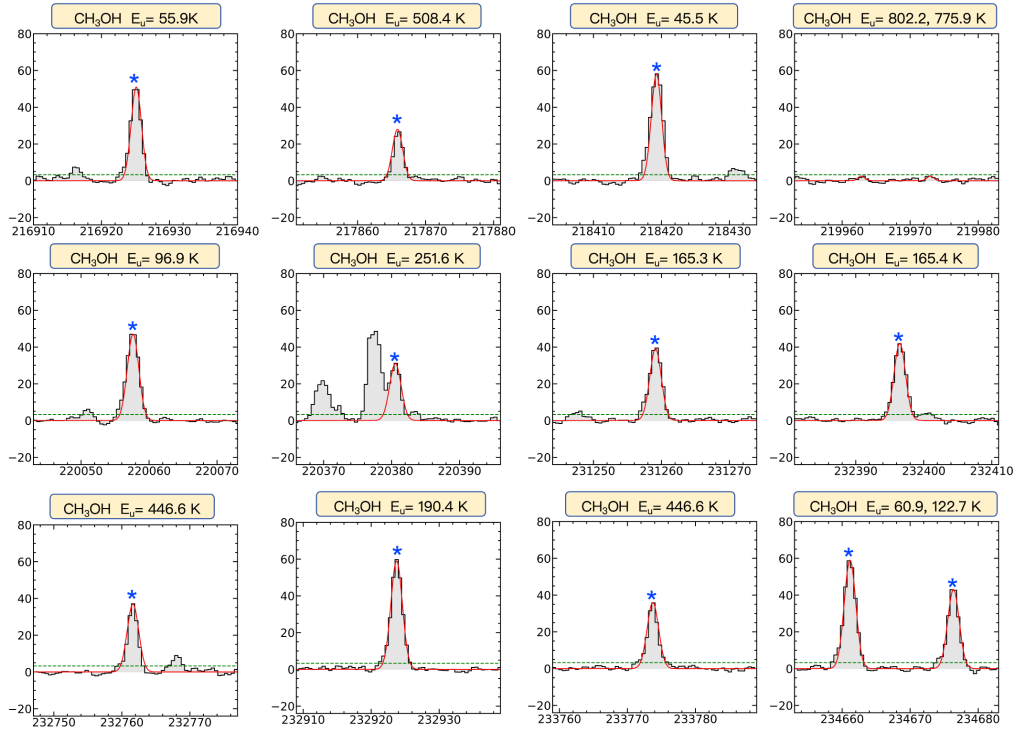}
   \caption{
Observed spectra and best-fit LTE models for selected spectral
windows containing CH$_3$OH transitions toward G35.2N MM3-pos. 
Black lines show the observed spectra, red lines represent the LTE models, 
and green dashed lines indicate the $3\sigma$ noise level, 
corresponding to approximately 3.3 K for an rms noise of $\sigma \simeq 1.1$ K. 
Blue asterisks mark the target spectral features used in the LTE analysis. 
The upper-state energies ($E_{\rm u}$) of the transitions are given above each panel.
}
\label{fig:CH3OH}
\end{figure}

\begin{figure}[!htp]
  \centering
  \includegraphics[width=0.50\textwidth]{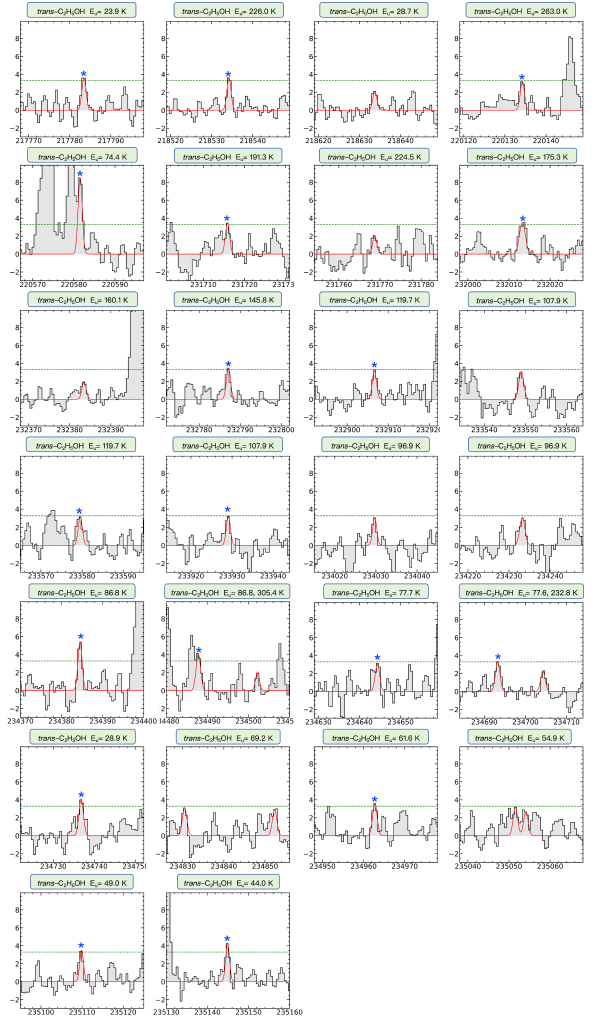}
  \caption{
Observed and best-fit LTE-modeled spectra of
\textit{trans}-C$_2$H$_5$OH toward G35.2N MM3-pos.
The observed spectra are shown in gray, and the LTE models are shown
in red. Other plotting conventions are the same as in
Fig.~\ref{fig:CH3OH}.
}
\label{fig:C2H5OH}
\end{figure}

\begin{figure}[!htp]
  \centering
  \includegraphics[width=0.50\textwidth]{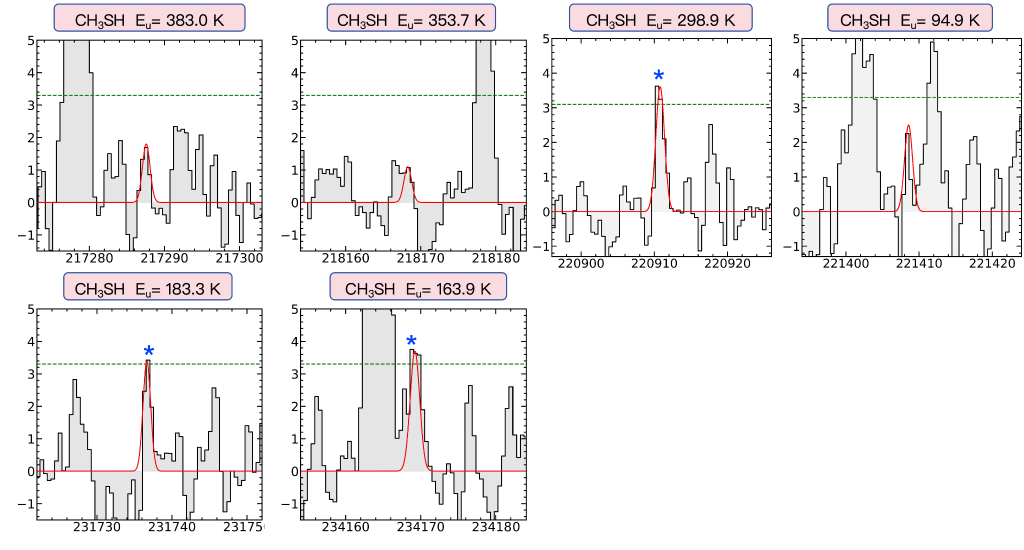} 
 \caption{
Observed and best-fit LTE-modeled spectra of CH$_3$SH toward
G35.2N MM3-pos. The observed spectra are shown in gray, and the
LTE models are shown in red. Other plotting conventions are the same
as in Fig.~\ref{fig:CH3OH}.
}
\label{fig:CH3SH}
\end{figure}

\subsection{Data reduction}

The data were calibrated and imaged using the Common
Astronomy Software Applications package (CASA; version
6.4.1-12; \citealt{McMullin2007}). The quasar J1924-2914 was
used for the bandpass and flux calibrations, while J1851+0035
was used for phase calibration. Line-free channels were selected
to construct the continuum data and to produce the
continuum-subtracted spectral-line data cubes. Phase-only
self-calibration was performed using the continuum data. The
shortest solution interval adopted for the phase self-calibration
was 6 s. The resulting self-calibration solutions were then
applied to the corresponding spectral-line data cubes.

The continuum images were produced with the CASA task
\texttt{tclean}, using Briggs weighting with a robust parameter
of 0.5. The continuum image shown in Fig.~\ref{fig:cont} has a
synthesized beam of $0\farcs29 \times 0\farcs28$, with a
position angle of $29.0\degr$. The achieved $1\sigma$ rms noise
levels are about 0.36\,mJy\,beam$^{-1}$ for the C-5 continuum
image and 0.27\,mJy\,beam$^{-1}$ for the continuum image shown
in Fig.~\ref{fig:cont}.

The spectral-line cubes used for the molecular analysis were
produced from the C-5 data for each spectral window. They
were imaged using Briggs weighting with a robust parameter of
0.5. We used the automatic masking procedure \texttt{YCLEAN}
\citep{Contreras2018}, which automatically cleans each channel
map with customized masks. More details on the \texttt{YCLEAN}
algorithm are given by \citet{Contreras2018}. The typical
$1\sigma$ rms noise level of the line cubes is about
10 mJy beam$^{-1}$ per channel. The largest recoverable angular
scale of the C-5 data is approximately $3\farcs0$.

\section{Results} \label{sec:res}

\subsection{Molecular identification}
\label{sec:res_spec}

The spectroscopic information for CH$_3$OH, CH$_3$SH,
C$_2$H$_5$OH, and C$_2$H$_5$SH was taken from the Cologne
Database for Molecular Spectroscopy (CDMS;
\citealt{Muller2001,Muller2005})\footnote{\url{https://cdms.astro.uni-koeln.de}}.
Only transitions from the ground vibrational state were considered.
The spectroscopic parameters of the transitions used for molecular
identification and modeling are listed in Appendix~\ref{app:A}. 
For C$_2$H$_5$OH, the \textit{trans} conformer, also referred to as
\textit{anti} in some spectroscopic catalogs, and the
\textit{gauche} conformer were inspected separately because they
have distinct rotational spectra. For C$_2$H$_5$SH, the
lower-energy \textit{gauche} conformer was modeled
\citep{Muller2016}.

Following the nomenclature defined in
Fig.~\ref{fig:cont}b, the quantitative local thermodynamic equilibrium (LTE) 
analysis was performed at MM3-pos, MM4-pos, and MM5-pos. These common
spectral-extraction positions were selected so that CH$_3$OH,
C$_2$H$_5$OH, CH$_3$SH, and C$_2$H$_5$SH could be examined
consistently at the same spatial locations with useful
signal-to-noise ratios and manageable line blending. They were not
selected as molecule-specific emission peaks and were not optimized
separately for individual species. For MM3 and MM4, 
where the adopted extraction positions are offset
from the corresponding continuum peaks, spectra were also extracted
at MM3-peak and MM4-peak for direct comparison. The continuum-peak
spectra were modeled independently using the same LTE procedure.
These fits were used only for the positional comparison, and the
resulting peak-position parameters were not included in the
column-density or abundance-ratio analysis reported below. 
The labels MM3, MM4, and MM5 are used here as spatial references to the corresponding
continuum structures. Because previous studies of G35.2N employed
different wavelengths, angular resolutions, and
component-identification schemes
\citep[e.g.,][]{Qiu2013,Sanchez2013,Sanchez2014,Beltran2016,
Zhang2022}, no one-to-one correspondence with earlier component
nomenclature is assumed.

\begin{table}
\centering
\footnotesize
\caption{LTE-based molecular parameters and adopted total column
densities at MM3-pos, MM4-pos, and MM5-pos.}
\label{tab:1}
\begin{tabular}{cccc}
\hline
Position & Species & $T_{\rm ex}$ & $N^{\rm f}_{\rm t}$\\
       &         &  (K) & (cm$^{-2}$) \\
\hline

\multirow{6}{*}{MM3-pos}
& CH$_3$OH$^{\rm a}$      & 130$\pm$30 & 1.3(17)$\pm$3(16) \\
& $^{13}$CH$_3$OH         & 120$\pm$20 & 3.6(16)$\pm$1.1(16) \\
& CH$_3$OH$^{\rm b}$      & --               & 1.8(18)$\pm$4(17) \\
& CH$_3$SH                & 110$\pm$30 & 1.7(16)$\pm$4(15) \\
& C$_2$H$_5$OH$^{\rm c}$  & 120$\pm$20 & 5.7(16)$\pm$1.1(16) \\
& C$_2$H$_5$SH$^{\rm d}$  & 110$\pm$20 & 3.3(15)$\pm$6(14) \\

\hline
\multirow{6}{*}{MM4-pos}
& CH$_3$OH$^{\rm a}$      & 130$\pm$20 & 1.5(17)$\pm$4(16) \\
& $^{13}$CH$_3$OH         & 110$\pm$20 & 4.0(16)$\pm$8(15) \\
& CH$_3$OH$^{\rm b}$      & --               & 2.0(18)$\pm$4(17) \\
& CH$_3$SH                & 100$\pm$20 & 1.9(16)$\pm$4(15) \\
& C$_2$H$_5$OH$^{\rm c}$  & 120$\pm$30 & 6.4(16)$\pm$8(15) \\
& C$_2$H$_5$SH$^{\rm d,e}$& 92$\pm$19        & 2.4(15)$\pm$4(14) \\

\hline
\multirow{6}{*}{MM5-pos}
& CH$_3$OH$^{\rm a}$      & 120$\pm$30 & 1.8(17)$\pm$4(16) \\
& $^{13}$CH$_3$OH         & 110$\pm$20 & 5.3(16)$\pm$1.1(16) \\
& CH$_3$OH$^{\rm b}$      & --               & 2.7(18)$\pm$5(17) \\
& CH$_3$SH                & 110$\pm$20 & 2.6(16)$\pm$5(15) \\
& C$_2$H$_5$OH$^{\rm c}$  & 120$\pm$20 & 8.5(16)$\pm$1.1(16) \\
& C$_2$H$_5$SH$^{\rm d}$  & 100$\pm$20 & 4.6(15)$\pm$9(14) \\

\hline
\end{tabular}

\tablefoot{
\scriptsize
$^{\rm a}$ Best-fit column density of the main CH$_3$OH isotopologue.
This value is affected by optical-depth effects and is not used in the
ratio analysis.\\
$^{\rm b}$ Adopted total CH$_3$OH column density derived from
$^{13}$CH$_3$OH using $^{12}$C/$^{13}$C = 50
\citep{Sanchez2014,Allen2017}.\\
$^{\rm c}$ Total C$_2$H$_5$OH column density derived from
\textit{trans}-C$_2$H$_5$OH transitions, using a partition function
that includes both conformers.\\
$^{\rm d}$ Total C$_2$H$_5$SH column density derived from
\textit{gauche}-C$_2$H$_5$SH transitions, using a partition function
that includes both conformers.\\
$^{\rm e}$ The MM4-pos value of C$_2$H$_5$SH is tentative
because the corresponding \textit{gauche}-C$_2$H$_5$SH features do
not meet the robust-identification criterion and are detected only at
the $\sim 2\sigma$ level.\\
$^{\rm f}$ X(Y) means $X \times 10^Y$.
}
\end{table}

The observed spectra were modeled under the assumption of LTE using
\texttt{CLASS/Weeds} \citep{Mar2011} within GILDAS, and the fits were
further refined with \texttt{CLASS/ADJUST}. Five parameters were
considered in the LTE modeling: source size ($\theta$), linewidth
($\Delta V$), source velocity ($V_{\rm LSR}$), excitation temperature
($T_{\rm ex}$), and column density ($N_{\rm t}$). To reduce parameter
degeneracy, only $T_{\rm ex}$ and $N_{\rm t}$ were treated as free
parameters, while $\theta$, $\Delta V$, and $V_{\rm LSR}$ were fixed
to the deconvolved continuum size, the average linewidth measured
from bright unblended lines, and 32.0~km~s$^{-1}$
\citep{Little1985}, respectively. These fixed parameters were adopted
consistently for all target species at a given extraction position,
providing a common basis for comparing the derived column densities
and abundance ratios.

Line identification was based on comparisons between the observed
spectra and synthetic spectra of the target species, together with
the modeled contributions of other molecules in the same spectral
windows. A molecular identification was classified as robust when at
least three spectrally independent target features had observed peak
intensities above the adopted $3\sigma$ threshold, were unblended or
only marginally blended, and were reproduced simultaneously by a
single LTE model. The line centers and profiles of these features had
to be compatible with the adopted $V_{\rm LSR}$ and linewidth, and
their relative intensities had to be reproduced by the model. A
spectrally independent feature refers to a distinct feature in the
observed spectrum rather than simply to an individual cataloged
transition. Unresolved spectroscopic components contributing to the
same observed feature were therefore counted only once. Features
below $3\sigma$ were not counted toward the robust-identification
criterion. In tentative cases, weak features were used only to place
provisional constraints on $N_{\rm t}$, whereas severely blended
features were excluded from both molecular identification and
quantitative fitting.

\begin{figure}[!htp]
  \centering
  \includegraphics[width=0.50\textwidth]{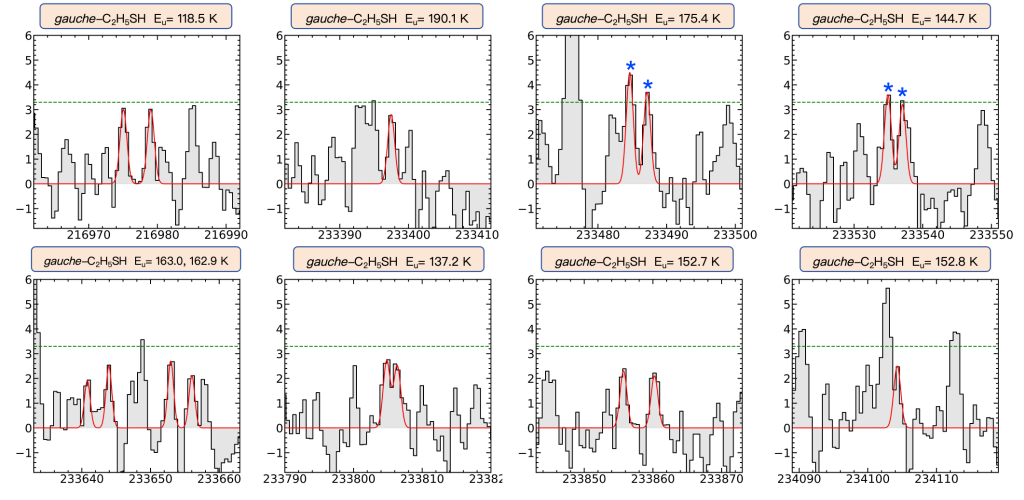} 
\caption{
Observed and best-fit LTE-modeled spectra of
\textit{gauche}-C$_2$H$_5$SH toward G35.2N MM3-pos.
The observed spectra are shown in gray, and the LTE models are shown
in red. Other plotting conventions are the same as in
Fig.~\ref{fig:CH3OH}.
}
\label{fig:C2H5SH}
\end{figure}

\begin{figure}[!htp]
  \centering
  \includegraphics[width=0.50\textwidth]{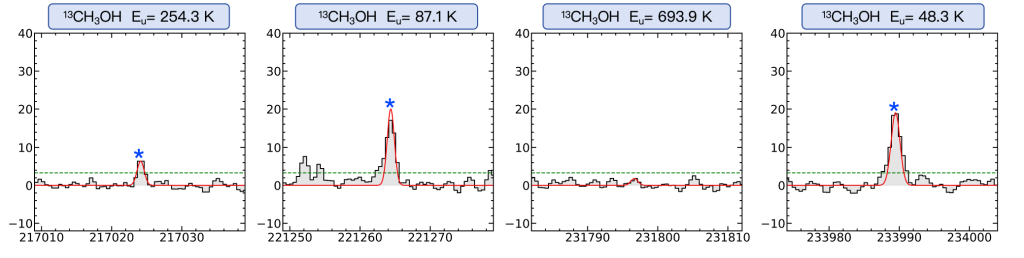}
\caption{
Observed and best-fit LTE-modeled spectra of
$^{13}$CH$_3$OH toward G35.2N MM3-pos.
The observed spectra are shown in gray, and the LTE models are shown
in red. Other plotting conventions are the same as in
Fig.~\ref{fig:CH3OH}.
}
\label{fig:13CH3OH}
\end{figure}

On this basis, CH$_3$OH, $^{13}$CH$_3$OH, CH$_3$SH, and
\textit{trans}-C$_2$H$_5$OH are robustly identified toward
MM3-pos, MM4-pos, and MM5-pos.
No robust identification of \textit{gauche}-C$_2$H$_5$OH was obtained.
The \textit{gauche}-C$_2$H$_5$SH identification is robust toward
MM3-pos and MM5-pos, whereas its emission toward
MM4-pos remains tentative at the $\sim2\sigma$ level and
provides only a provisional constraint on the column density.
Representative observed spectra and best-fit LTE models toward
MM3-pos are shown in Figs.~\ref{fig:CH3OH}--\ref{fig:13CH3OH}. 
The corresponding spectra toward MM4-pos and MM5-pos, 
together with those extracted toward MM3-peak and MM4-peak, 
are provided in the electronic supplementary material described in Appendix~\ref{app:B}. 
For each species, the same selected transitions, frequency windows, and panel
order are used at all five positions. Blue asterisks mark the target
features used in the LTE analysis at MM3-pos, MM4-pos, and MM5-pos.

The continuum-peak spectra show that CH$_3$SH and
\textit{gauche}-C$_2$H$_5$SH are more clearly identified toward
MM3-pos than toward MM3-peak. Toward MM4-peak, CH$_3$SH does not
satisfy the robust-identification criterion, and the
\textit{trans}-C$_2$H$_5$OH emission is weaker than toward
MM4-pos. The column densities and abundance ratios reported below
represent local measurements at MM3-pos, MM4-pos, and MM5-pos; they
are not intended to represent the corresponding continuum peaks,
core-averaged properties, or the complete spatial chemical variation
across G35.2N.

\begin{table}
\footnotesize
\centering
\caption{Abundance ratios for the alcohol/thioalcohol analog family.}
\label{tab:2}
\begin{tabular}{ccccc}
\hline
Position
& CH$_3$OH/
& CH$_3$OH/
& C$_2$H$_5$OH/
& CH$_3$SH/ \\
& C$_2$H$_5$OH
& CH$_3$SH
& C$_2$H$_5$SH
& C$_2$H$_5$SH \\

\hline

MM3-pos
& 32$\pm$9
& 110$\pm$30
& 17$\pm$5
& 5.2$\pm$1.4 \\

MM4-pos
& 31$\pm$8
& 110$\pm$30
& 27$\pm$5$^{\rm a}$
& 8$\pm$2$^{\rm a}$ \\

MM5-pos
& 32$\pm$8
& 100$\pm$30
& 19$\pm$4
& 5.7$\pm$1.6 \\

\hline
\end{tabular}

\tablefoot{
\scriptsize
The ratios were calculated from the adopted total column densities
listed in Table~\ref{tab:1}. $^{\rm a}$ Ratios marked with this
superscript use the tentative MM4-pos C$_2$H$_5$SH column
density listed in Table~\ref{tab:1}.
The ratios and their uncertainties were calculated from the
unrounded column densities and uncertainties; the rounded values are
reported in Table~\ref{tab:1}.
}
\end{table}

\subsection{Column densities and abundance ratios}
\label{sec:res_ratio}

The molecular parameters used in the abundance analysis are listed in
Table~\ref{tab:1}. To account for formal fitting uncertainties and
additional systematic effects, including residual line blending,
baseline placement, LTE assumptions, and partition-function
uncertainties, conservative uncertainties were adopted for the derived
$T_{\rm ex}$ and $N_{\rm t}$ values. The resulting relative
uncertainties are generally of order 20\%. The adopted total column
densities vary moderately across the three adopted extraction
positions. They range from $1.8\times10^{18}$ to
$2.7\times10^{18}$~cm$^{-2}$ for CH$_3$OH,
$1.7\times10^{16}$ to $2.6\times10^{16}$~cm$^{-2}$ for CH$_3$SH,
$5.7\times10^{16}$ to $8.5\times10^{16}$~cm$^{-2}$ for
C$_2$H$_5$OH, and $2.4\times10^{15}$ to
$4.6\times10^{15}$~cm$^{-2}$ for C$_2$H$_5$SH. The adopted
$^{12}$C/$^{13}$C ratio sets the absolute CH$_3$OH column-density
scale, and hence the absolute values of ratios involving CH$_3$OH,
but it does not affect the relative comparison among the three
selected positions because the same conversion factor is applied
throughout.

\begin{figure*}[!htp]
  \centering
  \includegraphics[width=0.98\textwidth]{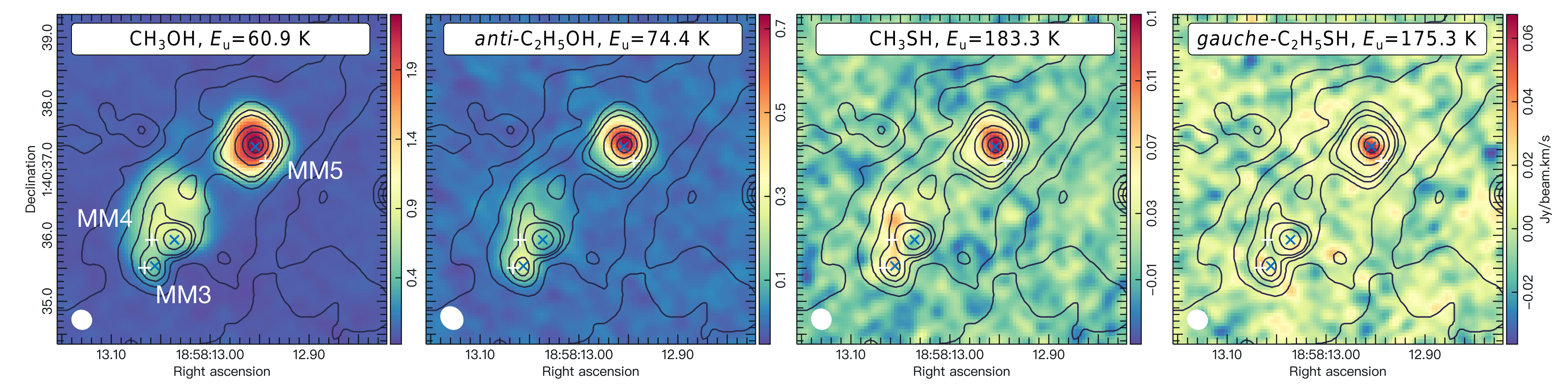}
  \caption{
Integrated-intensity maps of CH$_3$OH,
\textit{trans}-C$_2$H$_5$OH, CH$_3$SH, and
\textit{gauche}-C$_2$H$_5$SH toward the MM3--MM5 region of
G35.2N. The rms noise level of the integrated-intensity maps is
$2.8\times10^{-2}$ Jy beam$^{-1}$ km s$^{-1}$. The contours show
the 1.3 mm continuum emission from Fig.~\ref{fig:cont}, using the
same contour levels as in that figure. 
The ``$+$'' symbols mark MM3-pos, MM4-pos, and MM5-pos,
the spectral-extraction positions used in the quantitative LTE
analysis, whereas the ``$\times$'' symbols mark the corresponding
1.3~mm continuum peaks.
The white ellipses indicate the synthesized beam.
}
 \label{fig:4}
\end{figure*}

\begin{table*}[!htp]
\centering
\footnotesize
\caption{ALMA-based comparison samples for ratios within the alcohol/thioalcohol analog family.}
\label{tab:comparison_sample}
\begin{tabular}{llllllll}
\hline
Source & ALMA band & CH$_3$OH & C$_2$H$_5$OH & CH$_3$SH
& C$_2$H$_5$SH & Ref. & Ratio points$^{a}$ \\
\hline
G35.2N
& Band 6
& D
& D
& D
& D/P
& 1
& 3/3/3(1$^{p}$)/3(1$^{p}$) \\

Orion KL
& Band 6
& D
& D
& --
& --
& 2
& 3/0/0/0 \\

Sgr B2(N2)
& Band 3
& D
& D
& D$^{t}$
& UL
& 3
& 6/6/6$^{L}$/6$^{L}$ \\

G31.41+0.31 core
& Band 3
& D
& D
& D
& UL
& 4,5
& 1/1/1$^{L}$/1$^{L}$ \\

G31.41+0.31 shock
& Band 3
& D
& D
& D
& --
& 9
& 1/1/0/0 \\

IRAS 16293 B
& Band 7
& D
& D
& D
& UL
& 6--8
& 1/1/1$^{L}$/1$^{L}$ \\
\hline
\end{tabular}

\tablefoot{
\scriptsize
$^{a}$ The four numbers correspond to the number of available comparison
points for CH$_3$OH/C$_2$H$_5$OH, CH$_3$OH/CH$_3$SH,
C$_2$H$_5$OH/C$_2$H$_5$SH, and CH$_3$SH/C$_2$H$_5$SH,
respectively. D denotes detection, UL denotes upper limit, and ``--''
denotes not reported. For G35.2N, D/P indicates robust C$_2$H$_5$SH
constraints toward MM3 and MM5 and a tentative constraint toward MM4.
The notation 3(1$^{p}$) indicates three ratio points, one of which uses
the tentative MM4 C$_2$H$_5$SH column density. The superscript $L$
denotes ratios shown as lower limits because C$_2$H$_5$SH is constrained
by upper limits. The superscript $t$ denotes that one CH$_3$SH entry in
Sgr~B2(N2) is tentative. For the G31.41+0.31 shock, no
C$_2$H$_5$SH column density is available in the adopted comparison
study, and therefore only the CH$_3$OH/C$_2$H$_5$OH and
CH$_3$OH/CH$_3$SH ratios are included.
References: (1) this work;
(2)~\citet{Tercero2018}; (3)~\citet{Belloche2025};
(4)~\citet{Mininni2023}; (5)~\citet{Lopez2024};
(6)~\citet{Jorgensen2018}; (7)~\citet{Drozdovskaya2018};
(8)~\citet{Drozdovskaya2019}; (9)~\citet{Lopez2025}.
}
\end{table*}

The ratios listed in Table~\ref{tab:2} are used to compare the local
abundance relationships among the target alcohol/thioalcohol analogs.
We first consider CH$_3$OH/C$_2$H$_5$OH and CH$_3$OH/CH$_3$SH,
which provide the main ratio constraints in this work. Within the
adopted uncertainties, both ratios are consistent across
MM3-pos, MM4-pos, and MM5-pos, with nominal values of
31--32 and 100--110, respectively. 
The ratios involving C$_2$H$_5$SH span 17--27 for
C$_2$H$_5$OH/C$_2$H$_5$SH and 5.2--8 for
CH$_3$SH/C$_2$H$_5$SH. Their apparent spread is larger, but
the MM4-pos values remain tentative; these ratios are therefore
treated as secondary constraints.

\begin{table}[!htp]
\centering
\scriptsize
\caption{ALMA-based literature and modeled abundance ratios used for comparison.}
\label{tab:lit_model_ratios}
\setlength{\tabcolsep}{1.0pt}
\begin{tabular}{@{}llcccc@{}}
\hline
Source/model & Position/model
& CH$_3$OH/
& CH$_3$OH/
& C$_2$H$_5$OH/
& CH$_3$SH/ \\
 &
& C$_2$H$_5$OH
& CH$_3$SH
& C$_2$H$_5$SH
& C$_2$H$_5$SH \\
\hline

\multicolumn{6}{c}{ALMA-based literature ratios} \\

Orion KL & methyl formate (MF) peak
& 270 & -- & -- & -- \\

Orion KL & ethylene glycol (EG) peak
& 180 & -- & -- & -- \\

Orion KL & ethanol (ET) peak
& 51 & -- & -- & -- \\

Sgr B2(N2) & N2b
& 19 & 320 & $>$140 & $>$8.1 \\

Sgr B2(N2) & AN02 c1
& 21 & 130 & $>$35 & $>$5.8 \\

Sgr B2(N2) & AN02 c2
& 14 & 140 & $>$58 & $>$5.8 \\

Sgr B2(N2) & AN03
& 15 & 100 & $>$36 & $>$5.3 \\

Sgr B2(N2) & AN06c2
& 16 & 300$^{t}$ & $>$12 & $>$0.65$^{t}$ \\

Sgr B2(N2) & AN06
& 40 & 580 & $>$21 & $>$1.4 \\

G31.41+0.31 & core
& 170$\pm$40
& 1100$\pm$200
& $>$28
& $>$4.2 \\

G31.41+0.31 & shock
& 10$\pm$3
& 27$\pm$5
& --
& -- \\

IRAS 16293--2422 B & hot corino
& 43
& 2100
& $>$72
& $>$1.5 \\

\hline
\multicolumn{6}{c}{Warm-up model ratios} \\

Garrod22 & Model 1 (fast warm-up)
& 16.7 & 3500 & 3700 & 17.2 \\

Garrod22 & Model 2 (medium warm-up)
& 14.7 & 3200 & 3400 & 15.5 \\

Garrod22 & Model 3 (slow warm-up)
& 12.6 & 2200 & 2600 & 14.8 \\

\hline
\end{tabular}

\tablefoot{
\scriptsize
The G35.2N ratios are listed in Table~\ref{tab:2} and are not
repeated here. ``--'' denotes that the corresponding molecule
was not reported in the cited ALMA study. The symbol ``$>$''
denotes a lower limit resulting from an upper limit on the
C$_2$H$_5$SH column density. The superscript $t$ marks ratios
involving the tentative CH$_3$SH entry toward Sgr~B2(N2).
The observational ratios were recalculated from the molecular
column densities reported in the studies listed in
Table~\ref{tab:comparison_sample}. When uncertainties were
available for both column densities, ratio uncertainties were
propagated assuming independent errors and rounded following the
convention adopted in Tables~\ref{tab:1} and~\ref{tab:2}. When
one or both input uncertainties were unavailable, no formal ratio
uncertainty is reported. Lower limits are not assigned symmetric
uncertainties. The ALMA-based literature ratios are derived from
molecular column densities, whereas the warm-up model ratios are
derived from peak gas-phase fractional abundances in the models
of~\citet{Garrod2022}. Model~1, Model~2, and Model~3 correspond
to the fast, medium, and slow warm-up models, respectively.
}
\end{table}

\begin{figure*}[!ht]
\sidecaption
\includegraphics[width=12cm]{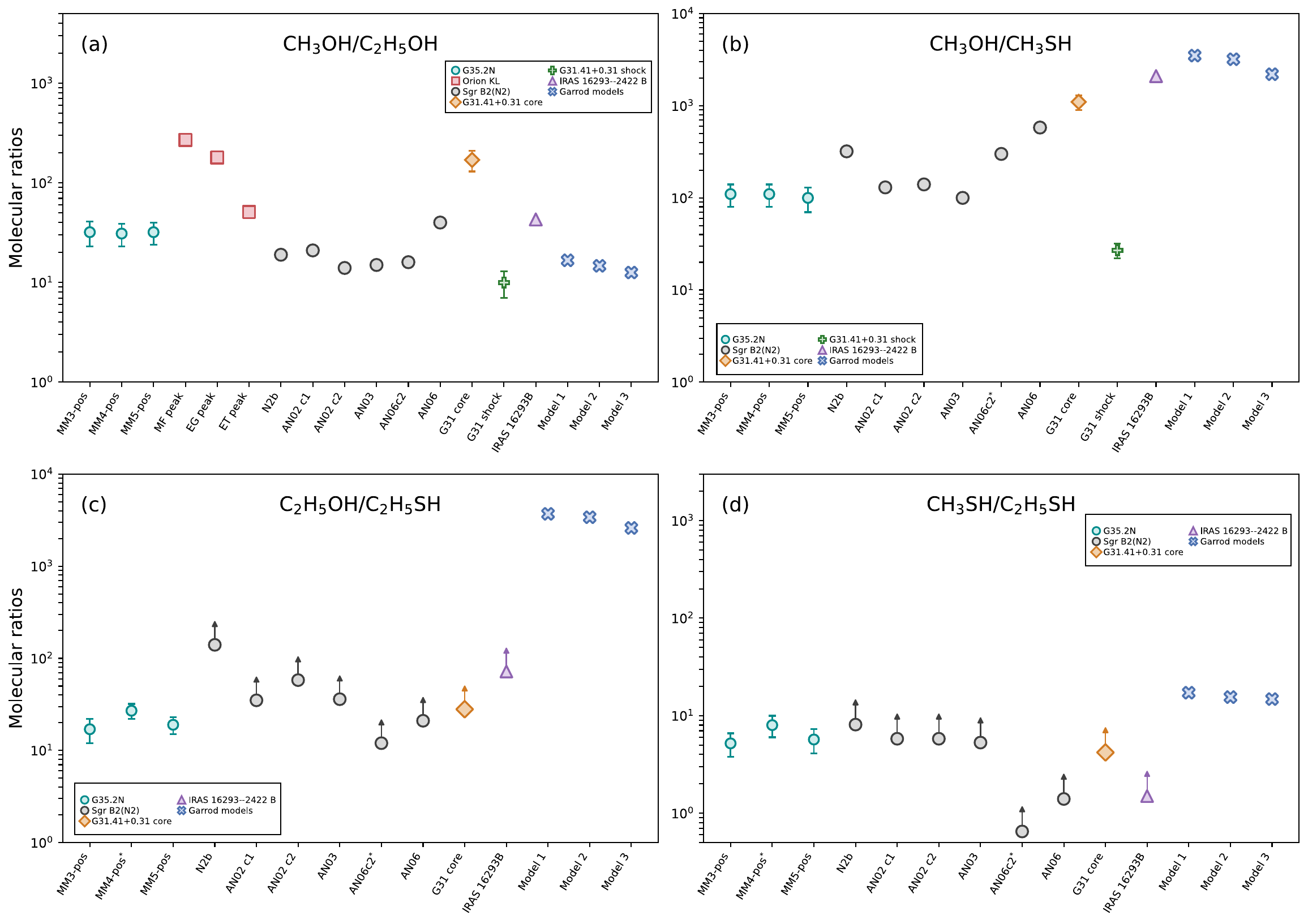}
\caption{Comparison of molecular abundance ratios within the
alcohol/thioalcohol analog family. Panels (a)--(d) show
CH$_3$OH/C$_2$H$_5$OH, CH$_3$OH/CH$_3$SH,
C$_2$H$_5$OH/C$_2$H$_5$SH, and
CH$_3$SH/C$_2$H$_5$SH, respectively. G35.2N values are shown
for MM3-pos, MM4-pos, and MM5-pos; the other source positions and
model designations are listed in Table~\ref{tab:lit_model_ratios}.
Error bars denote propagated observational uncertainties, upward
arrows denote lower limits, and asterisks mark ratios involving
tentative molecular constraints. The vertical axes are logarithmic.}
\label{fig:ratio_comparison}
\end{figure*}

\subsection{Spatial distributions and the nature of MM4-pos}
\label{sec:res_spatial}

Figure~\ref{fig:4} shows the integrated-intensity maps of CH$_3$OH,
\textit{trans}-C$_2$H$_5$OH, CH$_3$SH, and
\textit{gauche}-C$_2$H$_5$SH toward the MM3--MM5 region. The maps are
used only as qualitative spatial references because the selected
transitions have different strengths and signal-to-noise levels. Within
these limitations, the detected emission is broadly associated with the
compact 1.3 mm continuum structures.
CH$_3$OH and \textit{trans}-C$_2$H$_5$OH peak near the
northeastern MM5 continuum structure and extend southwestward across
the MM3--MM5 region. CH$_3$SH is likewise concentrated near MM5,
but its weaker emission limits a detailed morphological comparison.
The \textit{gauche}-C$_2$H$_5$SH map is noisier and is not used for
an independent morphological interpretation.

The offset between MM4-pos and the MM4 continuum peak is also
apparent in Fig.~\ref{fig:4} and in the comparison spectra in Appendix~\ref{app:B}.
The shock tracers were therefore examined to assess the nature of the molecular emission at MM4-pos.
The SiO $J = 5$ -- 4 transition is covered by the present spectral setup,
but its peak intensity toward MM4-pos remains below the adopted $3\sigma$ threshold.
The H$_2$CO and SO moment maps presented in Appendix~\ref{app:shock} show
extended emission and smoothly varying intensity-weighted velocities across the MM3--MM4 region,
without a distinct spatial or kinematic component centered on MM4-pos.
These observations do not provide sufficient evidence to classify MM4-pos as a separate shock position,
although a contribution from shock-affected or dynamically disturbed gas cannot be excluded.

\section{Discussion}
\label{sec:dis}

We compare the local abundance ratios in G35.2N with ALMA-based
measurements toward other chemically rich sources and with the
warm-up models of \citet{Garrod2022}. The comparison sample is
summarized in Tables~\ref{tab:comparison_sample} and
\ref{tab:lit_model_ratios} and shown in
Fig.~\ref{fig:ratio_comparison}. The comparison is not homogeneous:
the observational sample includes detections, lower limits, and a
small number of tentative entries, whereas the model ratios are based
on peak gas-phase fractional abundances rather than beam-averaged
column densities. We therefore use the comparison to characterize
broad abundance-ratio differences rather than to perform a
source-specific model fit. The multiplicative comparisons
quoted below are based on the nominal central values.

The CH$_3$OH/C$_2$H$_5$OH ratio ranges from 31 to 32 across
MM3-pos, MM4-pos, and MM5-pos, and the three values are
indistinguishable within their uncertainties. Across the observational
comparison sample, the ratio extends from $10\pm3$ at the
G31.41+0.31 shock position to 270 at the Orion~KL methyl-formate
peak, spanning a factor of about 30. The G35.2N values overlap the
range of 14--40 measured toward Sgr~B2(N2), are lower than the
G31.41+0.31 core value of $170\pm40$ by a factor of about five,
and are about three times the shock-position value. The nominal
core-to-shock contrast within G31.41+0.31 is a factor of about 20.
The warm-up models predict values of 12.6--16.7, approximately a
factor of two below the G35.2N measurements.

The CH$_3$OH/CH$_3$SH ratio is also consistent across
G35.2N, ranging from 100 to 110, but varies much more widely among
the comparison sources. The observational values extend from
$27\pm5$ at the G31.41+0.31 shock position to 2100 toward
IRAS~16293--2422~B, spanning a factor of about 80. The G35.2N
ratios are comparable to the lowest values measured toward
Sgr~B2(N2), where the ratio ranges from 100 to 580. The
G31.41+0.31 core value of $1100\pm200$ is about ten times higher
than the G35.2N values, whereas the G35.2N values are about four
times the shock-position value. The nominal core-to-shock contrast
within G31.41+0.31 is a factor of about 40. The warm-up model
predictions of 2200--3500 exceed the G35.2N values by nominal
factors of about 20--35.

The ratios involving C$_2$H$_5$SH provide less stringent
cross-source constraints because most literature values are lower
limits. In G35.2N, C$_2$H$_5$OH/C$_2$H$_5$SH ranges from 17 to
27, whereas the warm-up models predict 2600--3700, exceeding the
observed values by factors of about 100--200. The reported literature
lower bounds range from $>12$ to $>140$, but they do not establish
the relative ordering of the true ratios among sources. For
CH$_3$SH/C$_2$H$_5$SH, the G35.2N values of 5.2--8 are lower than
the model predictions of 14.8--17.2 by factors of about 2--3. The
corresponding literature constraints are also lower limits, with
reported bounds from $>0.65$ to $>8.1$. The MM4-pos ratios involving
C$_2$H$_5$SH remain provisional because the corresponding
\textit{gauche}-C$_2$H$_5$SH identification is tentative, and no
C$_2$H$_5$SH-based ratios are available for the G31.41+0.31 shock
position.

The two CH$_3$OH-based ratios are locally uniform across
MM3-pos, MM4-pos, and MM5-pos, but vary substantially among hot
cores, hot corinos, and shocked regions. By contrast, comparisons
involving C$_2$H$_5$SH remain limited by the prevalence of
nondetections. The observed differences may reflect variations in
grain desorption, subsequent gas-phase processing, evolutionary
conditions, or the available sulfur reservoir. Sulfur may remain
stored in refractory or icy material in dense star-forming regions
\citep{Laas2019,Shingledecker2020}, whereas shocks can release
sulfur-bearing material from grains into the gas phase
\citep{Requena2006,Zeng2018}. These ratios provide quantitative
observational constraints on sulfur-bearing organic chemistry,
although the available data do not distinguish between these
processes.

\section{Summary}
\label{sec:sum}

Using ALMA Band~6 data, we compared CH$_3$OH, CH$_3$SH,
C$_2$H$_5$OH, and C$_2$H$_5$SH toward
three selected spectral-extraction positions, MM3-pos,
MM4-pos, and MM5-pos, in the MM3--MM5 region of G35.2N.
LTE spectral modeling was used to derive column densities and local
abundance ratios. CH$_3$OH was constrained through
$^{13}$CH$_3$OH because the main isotopologue is optically thick.
CH$_3$OH, CH$_3$SH, and C$_2$H$_5$OH are robustly constrained at
all three positions, whereas
C$_2$H$_5$SH is robustly constrained at MM3-pos and MM5-pos
but remains tentative at MM4-pos, making the MM4-pos ethyl-level
ratios provisional.

The CH$_3$OH/C$_2$H$_5$OH and CH$_3$OH/CH$_3$SH ratios show no
measurable variation within the current uncertainties. The
CH$_3$OH/C$_2$H$_5$OH ratio is confined to 31--32, and the
CH$_3$OH/CH$_3$SH ratio varies only from 100 to 110,
indicating locally consistent CH$_3$OH/C$_2$H$_5$OH and
CH$_3$OH/CH$_3$SH abundance ratios across the selected positions.
The ratios involving C$_2$H$_5$SH span 17--27 for
C$_2$H$_5$OH/C$_2$H$_5$SH and 5.2--8 for
CH$_3$SH/C$_2$H$_5$SH, although the MM4-pos values are
tentative.

Comparison with ALMA-based literature measurements and warm-up
chemical models shows that the CH$_3$OH/C$_2$H$_5$OH ratio in
G35.2N lies within the range measured in other chemically rich
sources, whereas CH$_3$OH/CH$_3$SH exhibits a larger
source-to-source variation.
These results identify CH$_3$OH/CH$_3$SH as the best-constrained O/S
alcohol--thioalcohol analog ratio in the present data, while
more sensitive measurements of C$_2$H$_5$SH are needed to
establish the source-to-source behavior of the ethyl-level
C$_2$H$_5$OH/C$_2$H$_5$SH ratio.

\section*{Data availability}
\label{sec:sec6}

This work makes use of the following ALMA data:
ADS/JAO.ALMA~\#2021.1.00713.S.
The raw data are available from the ALMA Science Archive
(\url{https://almascience.nao.ac.jp/aq}) by entering the project
code listed above. ALMA is a partnership of ESO (representing its
member states), NSF (USA) and NINS (Japan), together with NRC
(Canada), NSTC and ASIAA (Taiwan), and KASI (Republic of Korea),
in cooperation with the Republic of Chile. The Joint ALMA
Observatory is operated by ESO, AUI/NRAO and NAOJ. The additional
LTE spectral fits and continuum-peak comparison spectra are
available at Zenodo
(\url{https://doi.org/10.5281/zenodo.20339046}).

\begin{acknowledgements}
We are grateful for support from the National SKA Program of China (No. 2025SKA0120100), National Natural Science Foundation of China (Grant No. W2512014), Fundamental Research Funds for the Central Universities (Grant No. 2025CDJ-IAISYB-060), and Postdoctoral Fellows Excellence Support Program (Grant No. 2404013554893087). D. L. is supported by the New Cornerstone Science Foundation.      
\end{acknowledgements}

\bibliographystyle{aa} 
\bibliography{aa61403-26} 

@ARTICLE{Santos2024,
       author = {{Santos}, Julia C. and {Enrique-Romero}, Joan and {Lamberts}, Thanja and {Linnartz}, Harold and {Chuang}, Ko-Ju},
        title = "{Formation of S-Bearing Complex Organic Molecules in Interstellar Clouds via Ice Reactions with C2H2, HS, and Atomic H}",
      journal = {ACS Earth and Space Chemistry},
         year = 2024,
        month = aug,
       volume = {8},
       number = {8},
        pages = {1646-1660},
          doi = {10.1021/acsearthspacechem.4c00150},
archivePrefix = {arXiv},
       eprint = {2407.09730},
 primaryClass = {astro-ph.GA},
       adsurl = {https://ui.adsabs.harvard.edu/abs/2024ESC.....8.1646S}
}

@ARTICLE{Lamberts2018,
       author = {{Lamberts}, T.},
        title = "{From interstellar carbon monosulfide to methyl mercaptan: paths of least resistance}",
      journal = {\aap},
         year = 2018,
        month = jul,
       volume = {615},
          eid = {L2},
        pages = {L2},
          doi = {10.1051/0004-6361/201832830},
archivePrefix = {arXiv},
       eprint = {1806.02990},
 primaryClass = {astro-ph.GA},
       adsurl = {https://ui.adsabs.harvard.edu/abs/2018A&A...615L...2L}
}

@ARTICLE{Muller2016,
       author = {{M{\"u}ller}, Holger S.~P. and {Belloche}, Arnaud and {Xu}, Li-Hong and {Lees}, Ronald M. and {Garrod}, Robin T. and {Walters}, Adam and {van Wijngaarden}, Jennifer and {Lewen}, Frank and {Schlemmer}, Stephan and {Menten}, Karl M.},
        title = "{Exploring molecular complexity with ALMA (EMoCA): Alkanethiols and alkanols in Sagittarius B2(N2)}",
      journal = {\aap},
         year = 2016,
        month = mar,
       volume = {587},
          eid = {A92},
        pages = {A92},
          doi = {10.1051/0004-6361/201527470},
archivePrefix = {arXiv},
       eprint = {1512.05301},
 primaryClass = {astro-ph.GA},
       adsurl = {https://ui.adsabs.harvard.edu/abs/2016A&A...587A..92M}
}

@ARTICLE{Majumdar2016,
       author = {{Majumdar}, L. and {Gratier}, P. and {Vidal}, T. and {Wakelam}, V. and {Loison}, J.-C. and {Hickson}, K.~M. and {Caux}, E.},
        title = "{Detection of CH$_{3}$SH in protostar IRAS 16293-2422}",
      journal = {\mnras},
         year = 2016,
        month = may,
       volume = {458},
       number = {2},
        pages = {1859-1865},
          doi = {10.1093/mnras/stw457},
archivePrefix = {arXiv},
       eprint = {1602.07619},
 primaryClass = {astro-ph.GA},
       adsurl = {https://ui.adsabs.harvard.edu/abs/2016MNRAS.458.1859M}
}

@ARTICLE{Laas2019,
       author = {{Laas}, Jacob C. and {Caselli}, Paola},
        title = "{Modeling sulfur depletion in interstellar clouds}",
      journal = {\aap},
         year = 2019,
        month = apr,
       volume = {624},
          eid = {A108},
        pages = {A108},
          doi = {10.1051/0004-6361/201834446},
archivePrefix = {arXiv},
       eprint = {1903.01232},
 primaryClass = {astro-ph.GA},
       adsurl = {https://ui.adsabs.harvard.edu/abs/2019A&A...624A.108L}
}

@ARTICLE{Gorai2021,
       author = {{Gorai}, Prasanta and {Das}, Ankan and {Shimonishi}, Takashi and {Sahu}, Dipen and {Mondal}, Suman Kumar and {Bhat}, Bratati and {Chakrabarti}, Sandip K.},
        title = "{Identification of Methyl Isocyanate and Other Complex Organic Molecules in a Hot Molecular Core, G31.41+0.31}",
      journal = {\apj},
         year = 2021,
        month = feb,
       volume = {907},
       number = {2},
          eid = {108},
        pages = {108},
          doi = {10.3847/1538-4357/abc9c4},
archivePrefix = {arXiv},
       eprint = {2011.02226},
 primaryClass = {astro-ph.GA},
       adsurl = {https://ui.adsabs.harvard.edu/abs/2021ApJ...907..108G}
}

@article{Ceccarelli2017,
  author  = {Ceccarelli, Cecilia and Caselli, Paola and Fontani, Francesco and others},
  title   = {Seeds Of Life In Space (SOLIS): The Organic Composition Diversity at 300--1000 au Scale in Solar-type Star-forming Regions},
  journal = {\apj},
  year    = {2017},
  volume  = {850},
  number  = {2},
  pages   = {176},
  adsurl  = {https://ui.adsabs.harvard.edu/abs/2017ApJ...850..176C}
}

@ARTICLE{Muller2001,
       author = {{M{\"u}ller}, H.~S.~P. and {Thorwirth}, S. and {Roth}, D.~A. and {Winnewisser}, G.},
        title = "{The Cologne Database for Molecular Spectroscopy, CDMS}",
      journal = {\aap},
         year = 2001,
        month = apr,
       volume = {370},
        pages = {L49-L52},
          doi = {10.1051/0004-6361:20010367},
       adsurl = {https://ui.adsabs.harvard.edu/abs/2001A&A...370L..49M}
}

@ARTICLE{Muller2005,
       author = {{M{\"u}ller}, Holger S.~P. and {Schl{\"o}der}, Frank and {Stutzki}, J{\"u}rgen and {Winnewisser}, Gisbert},
        title = "{The Cologne Database for Molecular Spectroscopy, CDMS: a useful tool for astronomers and spectroscopists}",
      journal = {Journal of Molecular Structure},
         year = 2005,
        month = may,
       volume = {742},
       number = {1-3},
        pages = {215-227},
          doi = {10.1016/j.molstruc.2005.01.027},
       adsurl = {https://ui.adsabs.harvard.edu/abs/2005JMoSt.742..215M}
}

@ARTICLE{Mar2011,
       author = {{Maret}, S. and {Hily-Blant}, P. and {Pety}, J. and {Bardeau}, S. and {Reynier}, E.},
        title = "{Weeds: a CLASS extension for the analysis of millimeter and sub-millimeter spectral surveys}",
      journal = {\aap},
         year = 2011,
        month = feb,
       volume = {526},
          eid = {A47},
        pages = {A47},
          doi = {10.1051/0004-6361/201015487},
archivePrefix = {arXiv},
       eprint = {1012.1747},
 primaryClass = {astro-ph.IM},
       adsurl = {https://ui.adsabs.harvard.edu/abs/2011A&A...526A..47M}
}

@ARTICLE{Li2025,
       author = {{Li}, Shanghuo and {Beuther}, Henrik and {Oliva}, Andr{\'e} and {Elbakyan}, Vardan G. and {Offner}, Stella S.~R. and {Kuiper}, Rolf and {Qiu}, Keping and {Lu}, Xing and {Sanhueza}, Patricio and {Chen}, Huei-Ru Vivien and {Zhang}, Qizhou and {Olguin}, Fernando A. and {Lee}, Chang Won and {Pudritz}, Ralph E. and {Kong}, Shuo and {Kuruwita}, Rajika L. and {Luo}, Qiuyi and {Liu}, Junhao},
        title = "{Detection of a septuple stellar system in formation via disk fragmentation}",
      journal = {Nature Astronomy},
         year = 2025,
        month = dec,
       volume = {9},
        pages = {1833-1844},
          doi = {10.1038/s41550-025-02682-9},
archivePrefix = {arXiv},
       eprint = {2509.06787},
 primaryClass = {astro-ph.GA},
       adsurl = {https://ui.adsabs.harvard.edu/abs/2025NatAs...9.1833L}
}

@ARTICLE{Rodriguez2021,
       author = {{Rodr{\'\i}guez-Almeida}, Lucas F. and {Jim{\'e}nez-Serra}, Izaskun and {Rivilla}, V{\'\i}ctor M. and {Mart{\'\i}n-Pintado}, Jes{\'u}s and {Zeng}, Shaoshan and {Tercero}, Bel{\'e}n and {de Vicente}, Pablo and {Colzi}, Laura and {Rico-Villas}, Fernando and {Mart{\'\i}n}, Sergio and {Requena-Torres}, Miguel A.},
        title = "{Thiols in the Interstellar Medium: First Detection of HC(O)SH and Confirmation of C$_{2}$H$_{5}$SH}",
      journal = {\apjl},
         year = 2021,
        month = may,
       volume = {912},
       number = {1},
          eid = {L11},
        pages = {L11},
          doi = {10.3847/2041-8213/abf7cb},
archivePrefix = {arXiv},
       eprint = {2104.08036},
 primaryClass = {astro-ph.GA},
       adsurl = {https://ui.adsabs.harvard.edu/abs/2021ApJ...912L..11R}
}

@ARTICLE{Garrod2022,
       author = {{Garrod}, Robin T. and {Jin}, Miwha and {Matis}, Kayla A. and {Jones}, Dylan and {Willis}, Eric R. and {Herbst}, Eric},
        title = "{Formation of Complex Organic Molecules in Hot Molecular Cores through Nondiffusive Grain-surface and Ice-mantle Chemistry}",
      journal = {\apjs},
         year = 2022,
        month = mar,
       volume = {259},
       number = {1},
          eid = {1},
        pages = {1},
          doi = {10.3847/1538-4365/ac3131},
archivePrefix = {arXiv},
       eprint = {2110.09743},
 primaryClass = {astro-ph.GA},
       adsurl = {https://ui.adsabs.harvard.edu/abs/2022ApJS..259....1G}
}

@ARTICLE{Shingledecker2020,
       author = {{Shingledecker}, Christopher N. and {Lamberts}, Thanja and {Laas}, Jacob C. and {Vasyunin}, Anton and {Herbst}, Eric and {K{\"a}stner}, Johannes and {Caselli}, Paola},
        title = "{Efficient Production of S$_{8}$ in Interstellar Ices: The Effects of Cosmic-Ray-driven Radiation Chemistry and Nondiffusive Bulk Reactions}",
      journal = {\apj},
         year = 2020,
        month = jan,
       volume = {888},
       number = {1},
          eid = {52},
        pages = {52},
          doi = {10.3847/1538-4357/ab5360},
archivePrefix = {arXiv},
       eprint = {1911.01239},
 primaryClass = {astro-ph.GA},
       adsurl = {https://ui.adsabs.harvard.edu/abs/2020ApJ...888...52S}
}

@ARTICLE{Requena2006,
       author = {{Requena-Torres}, M.~A. and {Mart{\'\i}n-Pintado}, J. and {Rodr{\'\i}guez-Franco}, A. and {Mart{\'\i}n}, S. and {Rodr{\'\i}guez-Fern{\'a}ndez}, N.~J. and {de Vicente}, P.},
        title = "{Organic molecules in the Galactic center. Hot core chemistry without hot cores}",
      journal = {\aap},
         year = 2006,
        month = sep,
       volume = {455},
       number = {3},
        pages = {971-985},
          doi = {10.1051/0004-6361:20065190},
archivePrefix = {arXiv},
       eprint = {astro-ph/0605031},
 primaryClass = {astro-ph},
       adsurl = {https://ui.adsabs.harvard.edu/abs/2006A&A...455..971R}
}

@ARTICLE{Zeng2018,
       author = {{Zeng}, S. and {Jim{\'e}nez-Serra}, I. and {Rivilla}, V.~M. and {Mart{\'\i}n}, S. and {Mart{\'\i}n-Pintado}, J. and {Requena-Torres}, M.~A. and {Armijos-Abenda{\~n}o}, J. and {Riquelme}, D. and {Aladro}, R.},
        title = "{Complex organic molecules in the Galactic Centre: the N-bearing family}",
      journal = {\mnras},
         year = 2018,
        month = aug,
       volume = {478},
       number = {3},
        pages = {2962-2975},
          doi = {10.1093/mnras/sty1174},
archivePrefix = {arXiv},
       eprint = {1804.11321},
 primaryClass = {astro-ph.GA},
       adsurl = {https://ui.adsabs.harvard.edu/abs/2018MNRAS.478.2962Z}
}

@INPROCEEDINGS{McMullin2007,
       author = {{McMullin}, J.~P. and {Waters}, B. and {Schiebel}, D. and {Young}, W. and {Golap}, K.},
        title = "{CASA Architecture and Applications}",
    booktitle = {Astronomical Data Analysis Software and Systems XVI},
         year = 2007,
       editor = {{Shaw}, R.~A. and {Hill}, F. and {Bell}, D.~J.},
       series = {Astronomical Society of the Pacific Conference Series},
       volume = {376},
        month = oct,
        pages = {127},
       adsurl = {https://ui.adsabs.harvard.edu/abs/2007ASPC..376..127M}
}

@ARTICLE{Contreras2018,
       author = {{Contreras}, Yanett and {Sanhueza}, Patricio and {Jackson}, James M. and {Guzm{\'a}n}, Andr{\'e}s E. and {Longmore}, Steven and {Garay}, Guido and {Zhang}, Qizhou and {Nguyễn-Lu'o'ng}, Quang and {Tatematsu}, Ken'ichi and {Nakamura}, Fumitaka and {Sakai}, Takeshi and {Ohashi}, Satoshi and {Liu}, Tie and {Saito}, Masao and {Gomez}, Laura and {Rathborne}, Jill and {Whitaker}, Scott},
        title = "{Infall Signatures in a Prestellar Core Embedded in the High-mass 70 {\ensuremath{\mu}}m Dark IRDC G331.372-00.116}",
      journal = {\apj},
         year = 2018,
        month = jul,
       volume = {861},
       number = {1},
          eid = {14},
        pages = {14},
          doi = {10.3847/1538-4357/aac2ec},
archivePrefix = {arXiv},
       eprint = {1805.01802},
 primaryClass = {astro-ph.GA},
       adsurl = {https://ui.adsabs.harvard.edu/abs/2018ApJ...861...14C}
}

@ARTICLE{Sanchez2014,
       author = {{S{\'a}nchez-Monge}, {\'A}. and {Beltr{\'a}n}, M.~T. and {Cesaroni}, R. and {Etoka}, S. and {Galli}, D. and {Kumar}, M.~S.~N. and {Moscadelli}, L. and {Stanke}, T. and {van der Tak}, F.~F.~S. and {Vig}, S. and {Walmsley}, C.~M. and {Wang}, K.-S. and {Zinnecker}, H. and {Elia}, D. and {Molinari}, S. and {Schisano}, E.},
        title = "{A necklace of dense cores in the high-mass star forming region G35.20-0.74 N: ALMA observations}",
      journal = {\aap},
         year = 2014,
        month = sep,
       volume = {569},
          eid = {A11},
        pages = {A11},
          doi = {10.1051/0004-6361/201424032},
archivePrefix = {arXiv},
       eprint = {1406.4081},
 primaryClass = {astro-ph.GA},
       adsurl = {https://ui.adsabs.harvard.edu/abs/2014A&A...569A..11S}
}

@ARTICLE{Allen2017,
       author = {{Allen}, V. and {van der Tak}, F.~F.~S. and {S{\'a}nchez-Monge}, {\'A}. and {Cesaroni}, R. and {Beltr{\'a}n}, M.~T.},
        title = "{Chemical segregation in hot cores with disk candidates. An investigation with ALMA}",
      journal = {\aap},
         year = 2017,
        month = jul,
       volume = {603},
          eid = {A133},
        pages = {A133},
          doi = {10.1051/0004-6361/201629118},
archivePrefix = {arXiv},
       eprint = {1705.06346},
 primaryClass = {astro-ph.SR},
       adsurl = {https://ui.adsabs.harvard.edu/abs/2017A&A...603A.133A}
}

@ARTICLE{Zhang2022,
       author = {{Zhang}, Yichen and {Tanaka}, Kei E.~I. and {Tan}, Jonathan C. and {Yang}, Yao-Lun and {Greco}, Eva and {Beltran}, Maria T. and {Sakai}, Nami and {De Buizer}, James M. and {Rosero}, Viviana and {Fedriani}, Rub{\'e}n and {Garay}, Guido},
        title = "{Massive Protostars in a Protocluster - A Multi-scale ALMA View of G35.20-0.74N.}",
      journal = {\apj},
         year = 2022,
        month = sep,
       volume = {936},
       number = {1},
          eid = {68},
        pages = {68},
          doi = {10.3847/1538-4357/ac847f},
archivePrefix = {arXiv},
       eprint = {2207.11320},
 primaryClass = {astro-ph.GA},
       adsurl = {https://ui.adsabs.harvard.edu/abs/2022ApJ...936...68Z}
}

@ARTICLE{Sanchez2013,
       author = {{S{\'a}nchez-Monge}, {\'A}. and {Cesaroni}, R. and {Beltr{\'a}n}, M.~T. and {Kumar}, M.~S.~N. and {Stanke}, T. and {Zinnecker}, H. and {Etoka}, S. and {Galli}, D. and {Hummel}, C.~A. and {Moscadelli}, L. and {Preibisch}, T. and {Ratzka}, T. and {van der Tak}, F.~F.~S. and {Vig}, S. and {Walmsley}, C.~M. and {Wang}, K.-S.},
        title = "{A candidate circumbinary Keplerian disk in G35.20-0.74 N: A study with ALMA}",
      journal = {\aap},
         year = 2013,
        month = apr,
       volume = {552},
          eid = {L10},
        pages = {L10},
          doi = {10.1051/0004-6361/201321134},
archivePrefix = {arXiv},
       eprint = {1303.4242},
 primaryClass = {astro-ph.GA},
       adsurl = {https://ui.adsabs.harvard.edu/abs/2013A&A...552L..10S}
}

@ARTICLE{Beltran2016,
       author = {{Beltr{\'a}n}, M.~T. and {Cesaroni}, R. and {Moscadelli}, L. and {S{\'a}nchez-Monge}, {\'A}. and {Hirota}, T. and {Kumar}, M.~S.~N.},
        title = "{Binary system and jet precession and expansion in G35.20-0.74N}",
      journal = {\aap},
         year = 2016,
        month = sep,
       volume = {593},
          eid = {A49},
        pages = {A49},
          doi = {10.1051/0004-6361/201628588},
archivePrefix = {arXiv},
       eprint = {1606.03943},
 primaryClass = {astro-ph.SR},
       adsurl = {https://ui.adsabs.harvard.edu/abs/2016A&A...593A..49B}
}

@ARTICLE{Qiu2013,
       author = {{Qiu}, Keping and {Zhang}, Qizhou and {Menten}, Karl M. and {Liu}, Hauyu B. and {Tang}, Ya-Wen},
        title = "{From Poloidal to Toroidal: Detection of a Well-ordered Magnetic Field in the High-mass Protocluster G35.2-0.74 N}",
      journal = {\apj},
         year = 2013,
        month = dec,
       volume = {779},
       number = {2},
          eid = {182},
        pages = {182},
          doi = {10.1088/0004-637X/779/2/182},
archivePrefix = {arXiv},
       eprint = {1311.0566},
 primaryClass = {astro-ph.GA},
       adsurl = {https://ui.adsabs.harvard.edu/abs/2013ApJ...779..182Q}
}

@ARTICLE{Zhang2025,
       author = {{Zhang}, Qizhou and {Liu}, Junhao and {Zeng}, Lingzhen and {Soler}, J.~D. and {Chen}, Huei-Ru Vivien and {Ching}, Tao-Chung and {Ho}, Paul T.~P. and {Girart}, Josep Miquel and {Koch}, Patrick M. and {Lai}, Shih-Ping and {Li}, Shanghuo and {Li}, Zhi-Yun and {Liu}, Hauyu Baobab and {Qiu}, Keping and {Rao}, Ramprasad},
        title = "{Impact of Gravity on Changing Magnetic Field Orientations in a Sample of Massive Protostellar Clusters Observed with ALMA}",
      journal = {\apj},
         year = 2025,
        month = oct,
       volume = {992},
       number = {1},
          eid = {103},
        pages = {103},
          doi = {10.3847/1538-4357/adfdcb},
archivePrefix = {arXiv},
       eprint = {2508.18538},
 primaryClass = {astro-ph.GA},
       adsurl = {https://ui.adsabs.harvard.edu/abs/2025ApJ...992..103Z}
}

@ARTICLE{Dent1989,
       author = {{Dent}, W.~R.~F. and {Sandell}, G. and {Duncan}, W.~D. and {Robson}, E.~I.},
        title = "{The structure of dust discs around G 35.2N, NGC 2071 and LkH-alpha 234.}",
      journal = {\mnras},
         year = 1989,
        month = jun,
       volume = {238},
        pages = {1497-1512},
          doi = {10.1093/mnras/238.4.1497},
       adsurl = {https://ui.adsabs.harvard.edu/abs/1989MNRAS.238.1497D}
}

@ARTICLE{Little1985,
       author = {{Little}, L.~T. and {Dent}, W.~R.~F. and {Heaton}, B. and {Davies}, S.~R. and {White}, G.~J.},
        title = "{The rotating interstellar disc in G 35.2-0.74.}",
      journal = {\mnras},
         year = 1985,
        month = nov,
       volume = {217},
        pages = {227-238},
          doi = {10.1093/mnras/217.2.227},
       adsurl = {https://ui.adsabs.harvard.edu/abs/1985MNRAS.217..227L}
}

@ARTICLE{Tercero2018,
       author = {{Tercero}, B. and {Cuadrado}, S. and {L{\'o}pez}, A. and {Brouillet}, N. and {Despois}, D. and {Cernicharo}, J.},
        title = "{Chemical segregation of complex organic O-bearing species in Orion KL}",
      journal = {\aap},
         year = 2018,
        month = nov,
       volume = {620},
          eid = {L6},
        pages = {L6},
          doi = {10.1051/0004-6361/201834417},
archivePrefix = {arXiv},
       eprint = {1811.08765},
 primaryClass = {astro-ph.GA},
       adsurl = {https://ui.adsabs.harvard.edu/abs/2018A&A...620L...6T}
}

@ARTICLE{Belloche2025,
       author = {{Belloche}, A. and {Garrod}, R.~T. and {M{\"u}ller}, H.~S.~P. and {Morin}, N.~J. and {Willis}, S.~A. and {Menten}, K.~M.},
        title = "{Re-exploring Molecular Complexity with ALMA: Insights into chemical differentiation from the molecular composition of hot cores in Sgr B2(N2)}",
      journal = {\aap},
         year = 2025,
        month = jun,
       volume = {698},
          eid = {A143},
        pages = {A143},
          doi = {10.1051/0004-6361/202554411},
archivePrefix = {arXiv},
       eprint = {2505.03262},
 primaryClass = {astro-ph.GA},
       adsurl = {https://ui.adsabs.harvard.edu/abs/2025A&A...698A.143B}
}

@ARTICLE{Mininni2023,
       author = {{Mininni}, C. and {Beltr{\'a}n}, M.~T. and {Colzi}, L. and {Rivilla}, V.~M. and {Fontani}, F. and {Lorenzani}, A. and {L{\'o}pez-Gallifa}, {\'A}. and {Viti}, S. and {S{\'a}nchez-Monge}, {\'A}. and {Schilke}, P. and {Testi}, L.},
        title = "{The GUAPOS project. III. Characterization of the O- and N-bearing complex organic molecules content and search for chemical differentiation}",
      journal = {\aap},
         year = 2023,
        month = sep,
       volume = {677},
          eid = {A15},
        pages = {A15},
          doi = {10.1051/0004-6361/202245277},
archivePrefix = {arXiv},
       eprint = {2306.13563},
 primaryClass = {astro-ph.GA},
       adsurl = {https://ui.adsabs.harvard.edu/abs/2023A&A...677A..15M}
}

@ARTICLE{Lopez2024,
       author = {{L{\'o}pez-Gallifa}, {\'A}. and {Rivilla}, V.~M. and {Beltr{\'a}n}, M.~T. and {Colzi}, L. and {Mininni}, C. and {S{\'a}nchez-Monge}, {\'A}. and {Fontani}, F. and {Viti}, S. and {Jim{\'e}nez-Serra}, I. and {Testi}, L. and {Cesaroni}, R. and {Lorenzani}, A.},
        title = "{The GUAPOS project - V: The chemical ingredients of a massive stellar protocluster in the making}",
      journal = {\mnras},
         year = 2024,
        month = apr,
       volume = {529},
       number = {4},
        pages = {3244-3283},
          doi = {10.1093/mnras/stae676},
archivePrefix = {arXiv},
       eprint = {2403.02191},
 primaryClass = {astro-ph.GA},
       adsurl = {https://ui.adsabs.harvard.edu/abs/2024MNRAS.529.3244L}
}

@ARTICLE{Jorgensen2018,
       author = {{J{\o}rgensen}, J.~K. and {M{\"u}ller}, H.~S.~P. and {Calcutt}, H. and {Coutens}, A. and {Drozdovskaya}, M.~N. and {{\"O}berg}, K.~I. and {Persson}, M.~V. and {Taquet}, V. and {van Dishoeck}, E.~F. and {Wampfler}, S.~F.},
        title = "{The ALMA-PILS survey: isotopic composition of oxygen-containing complex organic molecules toward IRAS 16293-2422B}",
      journal = {\aap},
         year = 2018,
        month = dec,
       volume = {620},
          eid = {A170},
        pages = {A170},
          doi = {10.1051/0004-6361/201731667},
archivePrefix = {arXiv},
       eprint = {1808.08753},
 primaryClass = {astro-ph.SR},
       adsurl = {https://ui.adsabs.harvard.edu/abs/2018A&A...620A.170J}
}

@ARTICLE{Drozdovskaya2018,
       author = {{Drozdovskaya}, Maria N. and {van Dishoeck}, Ewine F. and {J{\o}rgensen}, Jes K. and {Calmonte}, Ursina and {van der Wiel}, Matthijs H.~D. and {Coutens}, Audrey and {Calcutt}, Hannah and {M{\"u}ller}, Holger S.~P. and {Bjerkeli}, Per and {Persson}, Magnus V. and {Wampfler}, Susanne F. and {Altwegg}, Kathrin},
        title = "{The ALMA-PILS survey: the sulphur connection between protostars and comets: IRAS 16293-2422 B and 67P/Churyumov-Gerasimenko}",
      journal = {\mnras},
         year = 2018,
        month = jun,
       volume = {476},
       number = {4},
        pages = {4949-4964},
          doi = {10.1093/mnras/sty462},
archivePrefix = {arXiv},
       eprint = {1802.02977},
 primaryClass = {astro-ph.SR},
       adsurl = {https://ui.adsabs.harvard.edu/abs/2018MNRAS.476.4949D}
}

@ARTICLE{Drozdovskaya2019,
       author = {{Drozdovskaya}, Maria N. and {van Dishoeck}, Ewine F. and {Rubin}, Martin and {J{\o}rgensen}, Jes K. and {Altwegg}, Kathrin},
        title = "{Ingredients for solar-like systems: protostar IRAS 16293-2422 B versus comet 67P/Churyumov-Gerasimenko}",
      journal = {\mnras},
         year = 2019,
        month = nov,
       volume = {490},
       number = {1},
        pages = {50-79},
          doi = {10.1093/mnras/stz2430},
archivePrefix = {arXiv},
       eprint = {1908.11290},
 primaryClass = {astro-ph.SR},
       adsurl = {https://ui.adsabs.harvard.edu/abs/2019MNRAS.490...50D}
}

@ARTICLE{Hwang2026,
       author = {{Hwang}, Jihye and {Sanhueza}, Patricio and {Girart}, Josep Miquel and {Stephens}, Ian W. and {Beltr{\'a}n}, Maria T. and {Law}, Chi Yan and {Zhang}, Qizhou and {Liu}, Junhao and {Cort{\'e}s}, Paulo and {Olguin}, Fernando A. and {Koch}, Patrick M. and {Nakamura}, Fumitaka and {Saha}, Piyali and {Wang}, Jia-Wei and {Xu}, Fengwei and {Beuther}, Henrik and {Morii}, Kaho and {Fern{\'a}ndez L{\'o}pez}, Manuel and {Jiao}, Wenyu and {Kim}, Kee-Tae and {Li}, Shanghuo and {Zapata}, Luis A. and {Kim}, Jongsoo and {Choudhury}, Spandan and {Cheng}, Yu and {Pattle}, Kate and {Eswaraiah}, Chakali and {Sandhyarani}, Panigrahy and {Dewangan}, L.~K. and {Jadhav}, O.~R.},
        title = "{Magnetic Fields in Massive Star-forming Regions (MagMaR). VI. Magnetic Field Dragging in the Filamentary High-mass Star-forming Region G35.20─0.74N Due to Gravity}",
      journal = {\aj},
         year = 2026,
        month = jan,
       volume = {171},
       number = {1},
          eid = {50},
        pages = {50},
          doi = {10.3847/1538-3881/ae18c9},
archivePrefix = {arXiv},
       eprint = {2510.25078},
 primaryClass = {astro-ph.GA},
       adsurl = {https://ui.adsabs.harvard.edu/abs/2026AJ....171...50H}
}

@ARTICLE{Lopez2025,
       author = {{L{\'o}pez-Gallifa}, {\'A}. and {Rivilla}, V.~M. and {Beltr{\'a}n}, M.~T. and {Colzi}, L. and {Fontani}, F. and {S{\'a}nchez-Monge}, {\'A}. and {Mininni}, C. and {Cesaroni}, R. and {Jim{\'e}nez-Serra}, I. and {Viti}, S. and {Lorenzani}, A.},
        title = "{The GUAPOS project: VI. The chemical inventory of shocked gas}",
      journal = {\aap},
         year = 2025,
        month = dec,
       volume = {704},
          eid = {A288},
        pages = {A288},
          doi = {10.1051/0004-6361/202556837},
archivePrefix = {arXiv},
       eprint = {2509.16094},
 primaryClass = {astro-ph.GA},
       adsurl = {https://ui.adsabs.harvard.edu/abs/2025A&A...704A.288L}
}

\begin{appendix}
\section{Spectroscopic line lists}
\label{app:A}

The spectroscopic line lists used in this work were retrieved from
the Cologne Database for Molecular Spectroscopy (CDMS;
\citealt{Muller2001,Muller2005}). Tables~\ref{tab:A1} and
\ref{tab:A2} list the spectroscopic parameters of the unblended or
only minimally blended transitions used for molecular identification
and LTE modeling. The listed quantities include the rest frequency,
upper-state energy $E_{\rm u}$, Einstein coefficient $A_{ij}$, and
upper-state degeneracy $g_{\rm u}$. Only transitions from the ground
vibrational state were considered.

\begin{table}[H]
\centering
\caption{Spectroscopic parameters of the CH$_3$OH, $^{13}$CH$_3$OH,
CH$_3$SH, and \textit{gauche}-C$_2$H$_5$SH transitions used in this work.}
\label{tab:A1}
\fontsize{6.5}{6.8}\selectfont
\setlength{\tabcolsep}{1.1pt}
\renewcommand{\arraystretch}{0.72}
\begin{tabular}{llcccc}
\hline\hline
Molecule & Transition & Rest frequency & $E_{\rm u}$ & $A_{ij}$ & $g_{\rm u}$ \\
         &            & (MHz)          & (K)         & (s$^{-1}$) & \\
\hline
\multirow{14}{*}{CH$_3$OH}
& $5_{1,4,1}-4_{2,3,1}$        & 216945.521 & 55.9  & $1.21\times10^{-5}$ & 44  \\
& $20_{11,9,1}-20_{0,20,1}$    & 217886.504 & 508.4 & $3.38\times10^{-5}$ & 164 \\
& $4_{2,3,1}-3_{1,2,1}$        & 218440.063 & 45.5  & $4.69\times10^{-5}$ & 36  \\
& $25_{3,23,1}-24_{4,20,1}$    & 219983.675 & 802.2 & $2.04\times10^{-5}$ & 204 \\
& $23_{5,18,1}-22_{6,17,1}$    & 219993.658 & 775.9 & $1.74\times10^{-5}$ & 188 \\
& $8_{0,8,1}-7_{1,6,1}$        & 220078.561 & 96.6  & $2.52\times10^{-5}$ & 68  \\
& $10_{5,6,2}-11_{4,8,2}$      & 220401.317 & 251.6 & $1.12\times10^{-5}$ & 84  \\
& $10_{2,9,0}-9_{3,6,0}$       & 231281.100 & 165.3 & $1.83\times10^{-5}$ & 84  \\
& $10_{2,8,0}-9_{3,7,0}$       & 232418.521 & 165.4 & $1.87\times10^{-5}$ & 84  \\
& $18_{3,16,0}-17_{4,13,0}$    & 232783.446 & 446.5 & $2.17\times10^{-5}$ & 148 \\
& $10_{3,7,2}-11_{2,9,2}$      & 232945.797 & 190.4 & $2.13\times10^{-5}$ & 84  \\
& $18_{3,15,0}-17_{4,14,0}$    & 233795.666 & 446.6 & $2.20\times10^{-5}$ & 248 \\
& $4_{2,3,0}-5_{1,4,0}$        & 234683.370 & 60.9  & $1.87\times10^{-5}$ & 36  \\
& $5_{4,2,2}-6_{3,3,2}$        & 234698.519 & 122.7 & $6.34\times10^{-6}$ & 44  \\
\hline
\multirow{5}{*}{$^{13}$CH$_3$OH}
& $14_{1,13,-0}-13_{2,12,-0}$  & 217044.616 & 254.3 & $2.37\times10^{-5}$ & 29 \\
& $10_{2,8,+0}-9_{3,7,+0}$     & 217399.550 & 162.4 & $1.53\times10^{-5}$ & 21 \\
& $8_{-1,8,0}-7_{0,7,0}$       & 221285.241 & 87.1  & $3.72\times10^{-5}$ & 17 \\
& $22_{1,21,0}-22_{0,22,0}$    & 231818.380 & 593.9 & $3.87\times10^{-5}$ & 45 \\
& $5_{1,5,+0}-4_{1,4,+0}$      & 234011.580 & 48.3  & $5.27\times10^{-5}$ & 11 \\
\hline
\multirow{6}{*}{CH$_3$SH}
& $24_{2,22,1}-24_{1,23,1}$   & 217308.325 & 383.0 & $2.44\times10^{-5}$ & 49 \\
& $5_{1,4,1}-4_{0,4,0}$       & 218189.014 & 353.7 & $2.42\times10^{-5}$ & 47 \\
& $14_{1,14,0}-13_{1,13,0}$   & 220931.579 & 298.9 & $2.36\times10^{-5}$ & 43 \\
& $9_{-3,7,1}-10_{-2,9,1}$    & 221430.573 & 94.9  & $5.04\times10^{-6}$ & 19 \\
& $16_{2,14,1}-16_{1,15,1}$   & 231758.730 & 183.3 & $2.19\times10^{-5}$ & 33 \\
& $15_{2,13,1}-15_{1,14,1}$   & 234191.307 & 163.9 & $2.16\times10^{-5}$ & 31 \\
\hline
\multirow{21}{*}{\textit{gauche}-C$_2$H$_5$SH}
& $21_{2,19,1}-20_{2,18,1}$   & 216094.512 & 118.6 & $1.28\times10^{-4}$ & 43 \\
& $21_{2,19,0}-20_{2,18,0}$   & 216098.465 & 118.5 & $1.28\times10^{-4}$ & 43 \\
& $23_{7,16,1}-22_{7,15,1}$   & 233415.881 & 190.1 & $1.46\times10^{-4}$ & 47 \\
& $23_{7,17,1}-22_{7,16,1}$   & 233415.881 & 190.1 & $1.46\times10^{-4}$ & 47 \\
& $23_{7,16,0}-22_{7,15,0}$   & 233418.706 & 190.1 & $1.46\times10^{-4}$ & 47 \\
& $23_{7,17,0}-22_{7,16,0}$   & 233418.706 & 190.1 & $1.46\times10^{-4}$ & 47 \\
& $23_{6,17,1}-22_{6,16,1}$   & 233505.437 & 175.4 & $1.51\times10^{-4}$ & 47 \\
& $23_{6,18,1}-22_{6,17,1}$   & 233505.437 & 175.4 & $1.51\times10^{-4}$ & 47 \\
& $23_{6,17,0}-22_{6,16,0}$   & 233508.333 & 175.3 & $1.51\times10^{-4}$ & 47 \\
& $23_{6,18,0}-22_{6,17,0}$   & 233508.333 & 175.3 & $1.51\times10^{-4}$ & 47 \\
& $23_{3,21,1}-22_{3,20,1}$   & 233555.839 & 144.8 & $1.59\times10^{-4}$ & 47 \\
& $23_{3,21,0}-22_{3,20,0}$   & 233558.146 & 144.7 & $1.59\times10^{-4}$ & 47 \\
& $23_{5,19,1}-22_{5,18,1}$   & 233661.831 & 163.0 & $1.54\times10^{-4}$ & 47 \\
& $23_{5,19,0}-22_{5,18,0}$   & 233664.894 & 162.9 & $1.54\times10^{-4}$ & 47 \\
& $23_{5,18,1}-22_{5,17,1}$   & 233673.956 & 163.0 & $1.54\times10^{-4}$ & 47 \\
& $23_{5,18,0}-22_{5,17,0}$   & 233676.987 & 162.9 & $1.54\times10^{-4}$ & 47 \\
& $23_{1,22,1}-22_{1,21,1}$   & 233825.787 & 137.2 & $1.61\times10^{-4}$ & 47 \\
& $23_{1,22,0}-22_{1,21,0}$   & 233827.368 & 137.2 & $1.61\times10^{-4}$ & 47 \\
& $23_{4,20,0}-22_{4,19,0}$   & 233876.707 & 152.7 & $1.57\times10^{-4}$ & 47 \\
& $23_{4,20,1}-22_{4,19,1}$   & 233881.242 & 152.8 & $1.57\times10^{-4}$ & 47 \\
& $23_{4,19,0}-22_{4,18,0}$   & 234125.485 & 152.8 & $1.58\times10^{-4}$ & 47 \\
\hline
\end{tabular}
\end{table}

\begin{table}[!t]
\centering
\caption{Spectroscopic parameters of the \textit{trans}-C$_2$H$_5$OH
transitions used in this work.}
\label{tab:A2}
\fontsize{6.5}{6.8}\selectfont
\setlength{\tabcolsep}{1.5pt}
\renewcommand{\arraystretch}{0.82}
\begin{tabular}{lcccc}
\hline\hline
Transition & Rest frequency & $E_{\rm u}$ & $A_{ij}$ & $g_{\rm u}$ \\
           & (MHz)          & (K)         & (s$^{-1}$) & \\
\hline
$5_{3,3,2}-4_{2,2,2}$       & 217803.743 & 23.9  & $6.34\times10^{-5}$ & 11 \\
$5_{3,2,2}-4_{2,3,2}$       & 218461.233 & 23.9  & $6.38\times10^{-5}$ & 11 \\
$21_{5,16,2}-21_{4,17,2}$   & 218554.507 & 226.0 & $7.18\times10^{-5}$ & 43 \\
$7_{2,5,2}-6_{1,6,2}$       & 218654.089 & 28.7  & $2.69\times10^{-5}$ & 15 \\
$24_{3,22,2}-24_{2,23,2}$   & 220154.771 & 263.0 & $4.95\times10^{-5}$ & 49 \\
$13_{1,13,2}-12_{0,12,2}$   & 220601.928 & 74.4  & $1.02\times10^{-4}$ & 27 \\
$21_{5,17,2}-21_{4,18,2}$   & 231558.568 & 226.0 & $8.36\times10^{-5}$ & 43 \\
$20_{5,16,2}-20_{4,17,2}$   & 231560.921 & 208.2 & $8.27\times10^{-5}$ & 41 \\
$19_{5,15,2}-19_{4,16,2}$   & 231737.621 & 191.3 & $8.18\times10^{-5}$ & 39 \\
$22_{5,18,2}-22_{4,19,2}$   & 231790.057 & 224.5 & $8.46\times10^{-5}$ & 45 \\
$18_{5,14,2}-18_{4,15,2}$   & 232034.642 & 175.3 & $8.10\times10^{-5}$ & 37 \\
$15_{5,10,2}-15_{4,11,2}$   & 232075.845 & 132.3 & $7.73\times10^{-5}$ & 31 \\
$23_{5,19,2}-23_{4,20,2}$   & 232318.514 & 264.0 & $8.59\times10^{-5}$ & 47 \\
$17_{5,13,2}-17_{4,14,2}$   & 232404.841 & 160.1 & $8.02\times10^{-5}$ & 35 \\
$16_{5,12,2}-16_{4,13,2}$   & 232808.862 & 145.8 & $7.93\times10^{-5}$ & 33 \\
$14_{5,9,2}-14_{4,10,2}$    & 232928.526 & 119.7 & $7.65\times10^{-5}$ & 29 \\
$13_{5,8,2}-13_{4,9,2}$     & 233571.052 & 107.9 & $7.54\times10^{-5}$ & 27 \\
$14_{5,10,2}-14_{4,11,2}$   & 233601.554 & 119.7 & $7.71\times10^{-5}$ & 29 \\
$13_{5,9,2}-13_{4,10,2}$    & 233951.232 & 107.9 & $7.57\times10^{-5}$ & 27 \\
$12_{5,7,2}-12_{4,8,2}$     & 234051.149 & 96.9  & $7.38\times10^{-5}$ & 25 \\
$12_{5,8,2}-12_{4,9,2}$     & 234255.241 & 96.9  & $7.39\times10^{-5}$ & 25 \\
$11_{5,6,2}-11_{4,7,2}$     & 234406.433 & 86.8  & $7.17\times10^{-5}$ & 23 \\
$11_{5,7,2}-11_{4,8,2}$     & 234509.668 & 86.8  & $7.10\times10^{-5}$ & 23 \\
$25_{5,21,2}-25_{4,22,2}$   & 234523.904 & 305.4 & $8.92\times10^{-5}$ & 51 \\
$10_{5,5,2}-10_{4,6,2}$     & 234666.142 & 77.6  & $6.89\times10^{-5}$ & 21 \\
$10_{5,6,2}-10_{4,7,2}$     & 234714.783 & 77.6  & $6.90\times10^{-5}$ & 21 \\
$23_{1,22,2}-23_{0,23,2}$   & 234725.620 & 232.8 & $3.17\times10^{-5}$ & 47 \\
$6_{3,4,2}-5_{2,3,2}$       & 234758.520 & 28.9  & $7.09\times10^{-5}$ & 13 \\
$9_{5,4,2}-9_{4,5,2}$       & 234852.863 & 69.2  & $6.53\times10^{-5}$ & 19 \\
$9_{5,5,2}-9_{4,6,2}$       & 234873.874 & 69.2  & $6.54\times10^{-5}$ & 19 \\
$8_{5,3,2}-8_{4,4,2}$       & 234984.059 & 61.6  & $6.05\times10^{-5}$ & 17 \\
$8_{5,4,2}-8_{4,5,2}$       & 234992.193 & 61.6  & $6.05\times10^{-5}$ & 17 \\
$7_{5,2,2}-7_{4,3,2}$       & 235073.336 & 54.9  & $5.37\times10^{-5}$ & 15 \\
$7_{5,3,2}-7_{4,4,2}$       & 235076.061 & 54.9  & $5.37\times10^{-5}$ & 15 \\
$6_{5,1,2}-6_{4,2,2}$       & 235131.408 & 49.0  & $4.36\times10^{-5}$ & 13 \\
$6_{5,2,2}-6_{4,3,2}$       & 235132.155 & 49.0  & $4.36\times10^{-5}$ & 13 \\
$5_{5,0,2}-5_{4,1,2}$       & 235166.820 & 44.0  & $2.77\times10^{-5}$ & 11 \\
$5_{5,1,2}-5_{4,2,2}$       & 235166.969 & 44.0  & $2.77\times10^{-5}$ & 11 \\
\hline
\end{tabular}
\end{table}

\section{Additional spectral fits and continuum-peak spectra}
\label{app:B}

The supplementary spectral figures described in this appendix
are available through the Zenodo repository specified in the
Data availability section. They include the LTE fits toward
MM4-pos and MM5-pos and the spectra extracted at MM3-peak and
MM4-peak. The continuum-peak spectra were fitted independently
using the same LTE procedure as for the adopted extraction
positions. These independent peak-position fits are shown only
for comparison; their fitted parameters are not used to derive
the column densities or abundance ratios reported in
Tables~\ref{tab:1} and \ref{tab:2}. For each target species or
isotopologue, the same transitions, frequency windows, and panel
order are displayed toward MM3-pos, MM4-pos, MM5-pos, MM3-peak,
and MM4-peak. The plotting conventions follow those defined in
Fig.~\ref{fig:CH3OH}.

\section{Shock-tracer diagnostics toward the MM3--MM5 region}
\label{app:shock}

\begin{figure}[!b]
\centering

\includegraphics[width=0.98\linewidth]
{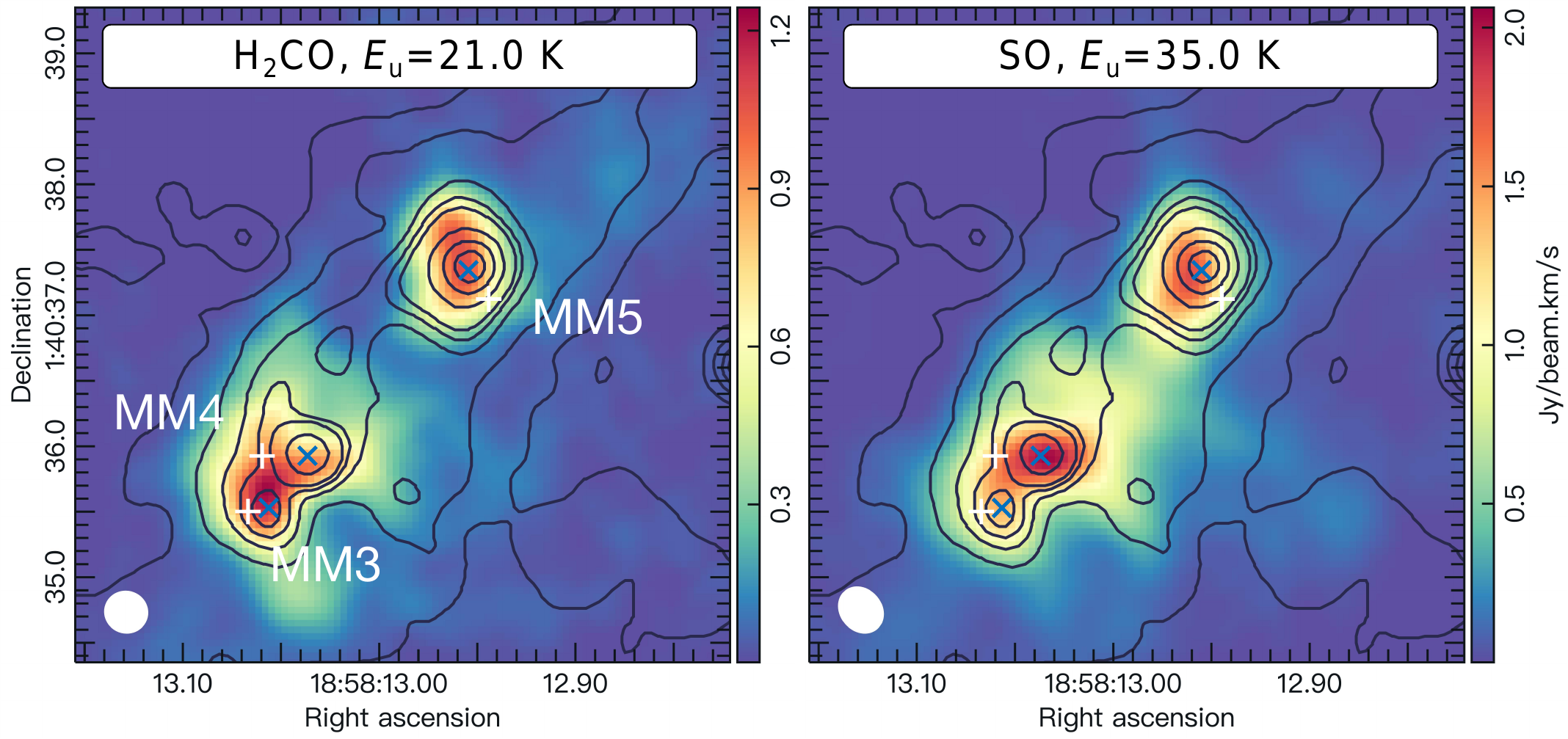}
\vspace{-0.5mm}

\includegraphics[width=0.98\linewidth]
{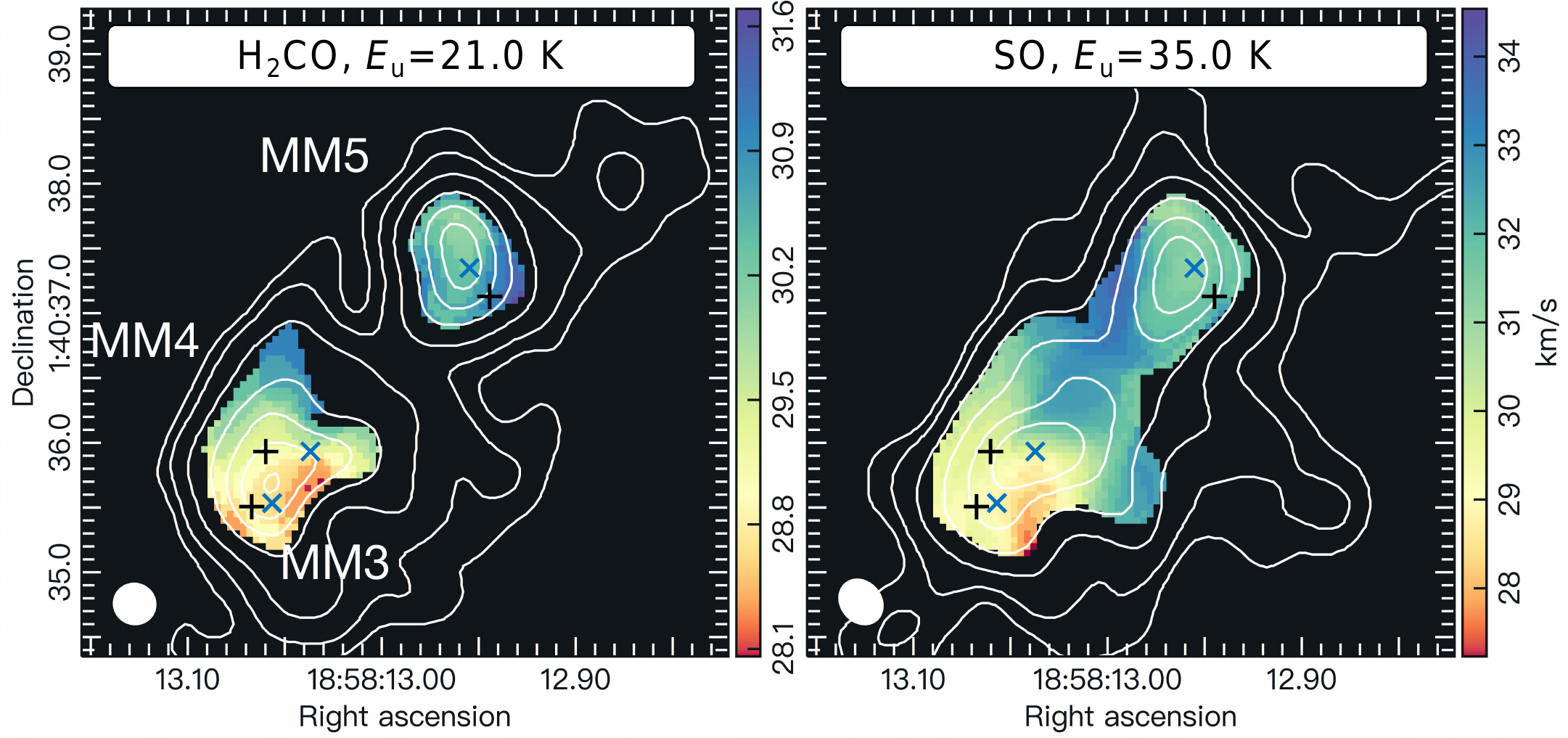}
\vspace{-2.5mm}

\caption{Moment-0 maps (upper panels) and moment-1 maps
(lower panels) of the H$_2$CO $3_{0,3}-2_{0,2}$ and SO
$6_5-5_4$ transitions toward the MM3--MM5 region. The
moment-0 contours show the 1.3~mm continuum emission, whereas
the moment-1 contours show the corresponding moment-0 emission.
The positional markers follow Fig.~\ref{fig:cont}b, and the white
ellipses indicate the synthesized beam. Only pixels above
$3\sigma$ were included in the moment-1 calculation.}
\label{fig:shock_maps}

\vspace{-2mm}
\end{figure}

The H$_2$CO and SO moment maps discussed in
Sect.~\ref{sec:res_spatial} are shown in Fig.~\ref{fig:shock_maps}.

\end{appendix}

\end{document}